# Physical Mechanism of Superconductivity

# Part 1 – High $T_c$ Superconductors

Xue-Shu Zhao, Yu-Ru Ge, Xin Zhao, Hong Zhao

ABSTRACT

The physical mechanism of superconductivity is proposed on the basis of carrier-induced dynamic strain effect. By this new model, superconducting state consists of the dynamic bound state of superconducting electrons, which is formed by the high-energy nonbonding electrons through dynamic interaction with their surrounding lattice to trap themselves into the three - dimensional potential wells lying in energy at above the Fermi level of the material. The binding energy of superconducting electrons dominates the superconducting transition temperature in the corresponding material. Under an electric field, superconducting electrons move coherently with lattice distortion wave and periodically exchange their excitation energy with chain lattice, that is, the superconducting electrons transfer periodically between their dynamic bound state and conducting state, so the superconducting electrons cannot be scattered by the chain lattice, and supercurrent persists in time. Thus, the intrinsic feature of superconductivity is to generate an oscillating current under a dc voltage. The wave length of an oscillating current equals the coherence length of superconducting electrons. The coherence lengths in cuprates must have the value equal to an even number times the lattice constant. A superconducting material must simultaneously satisfy the following three criteria required by superconductivity. First, the superconducting materials must possess high – energy nonbonding electrons with the certain concentrations required by their coherence lengths. Second, there must exist three – dimensional potential wells lying in energy at above the Fermi level of the material. Finally, the band structure of a superconducting material should have a widely dispersive antibonding band, which crosses the Fermi level and runs over the height of the potential wells to ensure the normal state of the material being metallic. According to the types of potential wells, the superconductors as a whole can be divided into two groups: the conventional and high temperature superconductors. The puzzling behavior of the cuprates, such as the complex phase diagrams, the linear dependence of resistivity with temperature in their normal states, the pseudogap, the transition temperature increasing with the number of the $CuO_2$ planes in the unit cell of Bi(Tl)-based compounds, the lattice instabilities and hardening in superconducting state, and the symmetries of superconducting waves, etc. all can be uniquely understood under this new model. In addition, the effects of strain and pressure, hole and electron doping, the replacement of trivalent rare-earth elements, and oxygen concentration on the superconducting properties of cuprates can be consistently explained by this physical mechanism. We demonstrate that the factor 2 in Josephson current equation, in fact, is resulting from 2V, the voltage drops across the two superconductor sections on both sides of a junction, not from the Cooper pair, and the magnetic flux is quantized in units of h/e, postulated by London, not in units of h/2e. The central features of superconductivity, such as Josephson effect, the tunneling mechanism in multijunction systems, and the origin of the superconducting tunneling phenomena, as well as the magnetic flux quantization in a superconducting hollow cylinder are all physically reconsidered under this superconductivity model. Following this unified superconductivity model, one will certainly know where to find the new materials with much higher $T_c$, even room temperatures superconductivity, and how to make high quality superconductor devices.

## I. INTRODUCTION

Since the discovery of superconductivity by Kammalingh Onner[1] in 1911, the study of superconductivity has always been a very active field both for the fundamental physics interest in the mystery of its origin and its many novel technical applications. However, the search for new materials with high transition temperature ($T_c$) has been mainly empirical, since no reliable theory is available for indicating the direction for finding high temperature superconductors. After a series of attempts to search for high-temperature superconductors, Matthais summarized some empirical rules for promoting transition temperature. One of them is that the materials with high transition temperature lie in the regime of lattice instabilities.[2] The superconductors found before 1986 usually are referred to as conventional ones in which the highest $T_c$ is 23.2 K occurred in $Nb_3Ge$ alloy with A-15 structure. It has been widely accepted for more than a half century that the BCS theory works well in the conventional superconductors.

The BCS microscopic theory was proposed by J. Bardeen, L Cooper, and T. R. Schrieffer in 1957 on the basis of Landau-Fermi liquid theory.[3] The main feature of the BCS theory may be summarized as follows: the two free electrons with opposite momentum and spin in the vicinity of Fermi surface form a bound electron pair, namely the Cooper pair, due to an exchange of virtual phonon. The Cooper pairs are allowed to have a large number overlaps between them. According to the BCS theory, superconductors at below $T_c$ have an energy gap equal to the binding energy of the Cooper pair, which dominates the transition temperature. The binding energy of the Cooper pair depends on the density of electron states at the Fermi surface, and on the strength of electron-phonon interaction. The pair states all have exactly the same net momentum, and scattering of an individual particle does not change the common momentum of the pair states, so the current persists in time.[4]

It is a general law of physics that the theory or model proposed for a certain physical subject must remain consistent with the entire physical structure. Since the BCS theory was proposed, it has never achieved a consistency with any field in condensed-matter physics. According to the basic concepts of quantum physics, it seems impossible that two free electrons by exchanging a phonon, an eigenstate of the lattice-harmonic vibrations, give rise to a lattice distortion which in turn leads to the Cooper pair bound state, since a lattice distortion must consist of a large number of phonons. In addition, where does the large amount of the lattice - distortion energy come from? If the lattice distortion is created by the electrostatic interaction between electrons and lattice, as the usual explanation, then the one electron of the Cooper pair must pay the lattice distortion energy by lowering its potential energy. Then it follows that this electron must become a deeply bound one with an energy state lying below the Fermi level, then turns out to be unable to carry any current. It is widely believed in solid state physics that a lattice distortion has a very short-range screened potential in metals (a couple of lattice constant), which in any case cannot become an attractive interaction source for the Cooper pairs with a coherence length as large as $10^2 – 10^4$ Å in conventional superconductors. However, in 1961, B. Deaver and F. William proposed that the magnetic flux threading a superconducting ring is quantized in units of h/2e (Ref. 5), and later in 1962, Josephson effect was discovered, which by chance confirmed that supercurrent tunnels through an insulating barrier by electron pairs.[6] Both effects imply that the supercurrent in superconductors is conducted by electron pairs. Since then, the Cooper pair concept unassailably underlies the microscopic theories of superconductivity.

Based on the theory of knowledge, the ability for scientists to cognize the laws of nature is limited by the development of science and technologies in their era. It is for this kind of reason that the correctness and generalization of the models or theories that we have accepted need to be continuously tested by the new discoveries. In 1986, the high temperature superconductivity in cuprate compounds was discovered by J. C. Bednorz and K. A. Müler, which raises an even crucial challenge to the BCS theory.[7] Because of the very short coherence length and strong electron-lattice interactions in cuprate compounds, most of the scientists strongly believe that the BCS theory cannot work in the High $T_c$ superconductors. Actually, it has been known for decades that organic and heavy fermion superconductors already cannot be well described by the BCS model.[8] After more than twenty years of intensive experimental and theoretical research, a great number of theoretical models have been proposed by using almost all of the elementary excitations of solids to mediate electron pairs. It is not surprising that none of them can be successful in explaining the basic feature of high $T_c$ superconductors. The real mechanism responsible for high $T_c$ superconductivity in cuprate compounds (in fact, for all superconductors) is still unclear, which has been acknowledged as one of the unsolved open questions in physics.

If a physical phenomenon keeps contradicting with any theory we have known, it perhaps means that there must be something missed in the fundamentals of physics. In other words, if a physical phenomenon was discovered so early as at that time the concept related to it had not existed in the fundamental physics, then no matter how hard people worked on it, the phenomenon could never be truly understood. What we have missed in condensed-matter physics for understanding the nature of superconductivity perhaps is the dynamic strain effect induced by the high - energy nonbonding electrons in open – shell compounds. In order to keep consistence with what we have found in nanocrystal field,[9,10] here we still call this effect as CIDSE (carrier - induced dynamic strain effect). The difference is that the CIDSE in nanocrystal systems is caused by the high-energy electrons excited by an external energy source, while the CIDSE in superconductors is resulting from the intrinsic high-energy nonbonding electrons of the material. While both of them can induce a dynamic bound state in the potential well lying in energy at above the Fermi level. We have found that CIDSE is a quite basic concept of condensed-matter physics, which plays an extremely important role in the understanding of the electronic and optical properties of the nano-size materials.[9,10]

We propose here that it is CIDSE that causes the superconductivity as long as the material simultaneously satisfies the following three necessary conditions. First, the material must contain the high-energy nonbonding electrons with a certain concentration required by the coherence length. Second, there must exist three-dimensional potential wells lying in energy at above the Fermi level of the material. Finally, the band structure of the superconducting material should have a widely dispersive antibonding band, which crosses the Fermi level and runs over the height of potential wells to ensure the normal state of the material being metallic.

In this new model, the superconducting state consists of the dynamic bound state of superconducting electrons, which is formed by the high-energy nonbonding electrons through dynamic interaction with their surrounding lattice to trap themselves into the three-dimensional potential wells lying in energy at above the Fermi level of the material. The binding energy of superconducting electrons dominates the superconducting transition temperature of the material. Under an electric field, the superconducting electrons move coherently with lattice distortion wave and periodically exchange their excitation energy with chain lattice, and so the superconducting electrons transfer periodically between their dynamic bound state and conducting state. So the superconducting electrons cannot be scattered by the chain lattice and the supercurrent persists without losing energy. Thus, the intrinsic feature of the superconductivity is to generate an oscillating supercurrent under a dc voltage. The wave length of an oscillating supercurrent equals the coherence length of the superconducting electrons. The coherence length of the superconducting electrons in cuprates must have the values equal to an even number times lattice constant. If the superconducting electrons have coherence length equal to an odd number times the lattice constant, then the superconducting process cannot be started. The superconducting state of a material must be accompanied by lattice distortion, and both the normal state and superconducting state for a given material must occur in the same crystalline phase no matter how large the lattice distortion could be. According to the types of potential wells into which the superconducting electrons trap themselves to form a superconducting dynamic bound state, the superconductors as a whole can be divided into two groups. One of them is called as usual as conventional superconductors, in which the three-dimensional potential wells are formed by the microstructures of materials, such as crystal grains, superlattice, nanocrystals, etc. The other one is referred to as high-$T_c$ superconductors in which the three-dimensional potential well is uniquely formed by the lattice structure of materials, like $CuO_6$ octahedrons, and $CuO_5$ pyramids in cuprates, and $C_{60}$ in $A_3C_{60}$ fullerrides, etc. In addition, almost all of the puzzling behavior of the cuprates observed from the point of view of the old theories, such as the complex phase diagram, the linear dependence of resistivity with temperature in their normal state, pseudogap, the symmetries of superconducting waves, and the $T_c$ increasing with the number of $CuO_2$ planes in the unit cell of Bi(Tl) - based compound, etc., all become the natural results of this new model. The correctness of this new model also lie in the following facts: the effects of hole and electron doping, the replacement of the trivalent rare - earth elements, oxygen concentration, strain and pressure on the superconducting properties of cuprates, all can be consistently explained by this new superconductivity mechanism. We demonstrate that the factor of 2 in Josephson equation, in fact, is resulting from 2V, the voltage drops across the two superconductor sections on both sides of a junction, not the Cooper pair, and the magnetic flux is quantized in units h/e, postulated by London, not in units of h/2e. Finally, the central features of superconductivity, like the Josephson effect, the superconducting tunneling mechanism in multijunction systems, and the origin of superconducting tunneling phenomena are all physically reconsidered under this unified superconductivity model.

However, the concept of CIDSE upon which the unified model of superconductivity can be built, has been missed in condensed-matter physics, so in the following section ( Sec. II ), we will systematically discuss the carrier-induced dynamic strain effect and some other concepts involved in this new model. Then on the basis of CIDSE, the unified physical mechanism of superconductivity is proposed in terms of two typical high – $T_c$ superconductors, $Ba_{1-x} K_x BiO_3$ and $La_{2-x}M_xCuO_4$ (in Sec.III). In Sec. IV, the central features of superconductivity, such as Josephson effect, the tunneling mechanism in multijunction systems, the origin of the superconducting tunneling phenomena, and the units of the magnetic flux quantization in a superconducting hollow cylinder are all physically reconsidered under this unified superconductivity model. In addition, the necessary conditions for achieving high quality superconductor devices are also discussed. It is concluded in Sec. V that all superconducting materials, no matter what the transition temperature is, must satisfy the three necessary conditions mentioned above, and the room temperatures superconductivity must lie in the materials in which the three criteria for superconductivity must be optimally satisfied.

## II. CARRIER-INDUCED DYNAMIC STRAIN EFFECT

It is widely believed that the first milestone for understanding the superconductivity was the London equation,[11] which successfully explained the Meissner effect. The London equation was derived on the basis of classic electromagnetic theory and two- fluid model. The two- fluid model assumed that only a fraction $n_s/n$ of total number of conduction electrons is able to carry the supercurrent in a superconductor at below transition temperature. The $n_s$ and n is the density of superconducting electrons and the total free electrons, respectively. The remaining fraction $n - n_s$ participates in the normal current. The normal current and supercurrent are assumed to flow in parallel. Since the supercurrent flows without any resistance, so the supercurrent effectively short-circuits the normal one, and carries all the current through a superconductor. Under an electric field, the supercurrent density and magnetic field have the following relation

$$\partial / \partial t [ \nabla \times ( J_s + n_s e^{*2} A / m^* )] = 0, \qquad B = \nabla \times A \qquad (1)$$

F. London and H. London discovered that in order to lead directly to the Meissiner effect, the equations above must have

$$J_s + n_s e^{*2} A/m^* = 0 \qquad (2)$$

This was known as the London equation. Where $J_s = v_s n_s e^*$, and $e^*$, $m^*$, and A is electronic charge, effective mass of electron, and vector potential of magnetic field B, respectively. From this equation, London demonstrated that supercurrent and magnetic field in superconductors can only exist within a surface layer of thickness, known as the London penetration depth, that is, the Meissner effect. Thus we can see that the London equation does hold the key feature of superconductivity. Although the London equation was derived from macroscopical theory, its result should predict some microscopical features that the superconducting electrons must follow. In a magnetic field, the drift velocity of a superconducting electron, $v_s$, is related to the canonical momentum P by

$$P = m^* v_s + (-e^* A) \qquad (3)$$

If we substitute the expression for A into the London equation above, it will be found that in order to keep both sides of the equation equal, the total momentum P of the superconducting electron must be zero.[12] Thus we can see that London equation provides a physical foundation for the assumption of the Cooper pair concept. For this reason, the BCS theory does explain some features of superconductivity.

In fact, we have another way to satisfy the requirement of the London equation that is to consider the superconducting electrons in bound states. The concept of bound state or localized state is introduced in condensed-matter physics as distinct from the free electron state. Generally, there are two kinds of bound states in the condensed-matter physics, one of them is the shallow bound state induced by the long-range Coulomb potential, which play an extremely important role in achieving the semiconductor devices, and the other one is the deep bound state created by the short-range potential, such as vacancies and the lattice distortion around a defect, etc. Note that by definition, both kinds of bound states are formed by the normal electron through electrostatic interaction with its surrounding lattice. In this sense, the Cooper pair should belong to the regime of the deep bound states, if it really exists. Since both kinds of bound states are located in energy below the Fermi level, so it appears that even if these two sorts of bound states do exist in superconductors, they definitely cannot participate to conduct any ballistic supercurrent. Thereby, we arrive at a conclusion that both kinds of bound states or localized states created by lowering the electron potential in the vicinity of a defect cannot match the requirement of the London equation. Fortunately, we have enough evidence to assume that in superconductors there exists another kind of bound state which is created by the high-energy nonbondling electron through dynamic interaction with its surrounding lattice to trap itself in a three - dimensional potential well lying in energy at above the Fermi level. It is just this kind of dynamic bound state that can match the requirement of the London equation and become the origin of superconductivity.

It has long been known that superconductivity usually occurs in open-shell compounds, such as s-p metals, transition metal compounds, and the alloy metals synthesized from the neighbor groups in the periodic table, as well as the copper oxide compounds, etc. In contrast to the covalent and closed-shell compound, there always exist the high-energy nonbonding electrons in superconducting materials. In order to show this feature, let us imagine that if we brought the atoms of an open-shell compound closer to form a solid, we would find that the large electron orbitals of s and p electrons at outer shells would first form the bonding bonds to hold the atoms together. While, the electrons at the d and f shells which are confined more closely to the nucleus than are the s and p states, do not strongly overlap with that of the neighboring atoms,[13] as in the cases of La 5d electrons in $La_2CuO_4$ and Y 4d electrons in $YBa_2Cu_3O_7$.

At high temperature these nonbonding high-energy electrons may still stay in their atomic orbitals. However, at the low temperature and equilibrium condition, it is energetically favorable for these high-energy nonbonding d - electrons to transfer to the available low - lying levels. In principle, the total energy of a solid should simply be the sum of energies carried by all individual valence electrons. Clearly, the energy carried by these nonbonding high - energy electrons must be a portion of the intrinsic total energy of a solid. Thus the transition energy released by the high-energy nonbonding electrons cannot be transferred out of the solid by emitting photons (electromagnetic energy) or phonons (thermal energy). The transition energy released by the high - energy nonbonding electrons must be transferred into other kinds of internal energies, such as to create lattice distortion, phase transition, or anharmonic off-site vibrations, or lattice dynamic strain in potential wells, etc. It is for this kind of reason that for almost all of superconductors, no matter how carefully to control the techniques which are required to produce single crystal, one still cannot obtain large perfect crystals. It seems to be a common feature that the higher the transition temperature, the smaller the size of crystal that one can obtain.

However, in modern condensed-matter physics, the conventional method for calculating the Fermi level of a given material is to integrate the density of states of the material to give the total number of valence electrons contained in a unit volume. That is, the energy levels in a superconductor are filled by the valence electrons from the lowest one up to the Fermi level. Once the Fermi level is found, the electronic and optical properties of a given superconductor can be uniquely studied in terms of the electronic density of states and electron-phonon interaction at the vicinity of the Fermi level, just like what the BCS theory did, because most of the physical properties of a solid can be determined by the electronic states around the Fermi level of the material. Obviously, during this treatment, the energies released by the nonbonding electrons transitions are totally ignored in the fundamentals of condensed-matter physics. For this reason, all physical phenomena related to this kind of energy, such as lattice distortions, phase transitions, anharmonic off-site vibrations and superconductivity can never be truly understood in the old theories.

Now let us consider the effect of these high-energy nonbonding electrons on the physical properties of a solid. From the thermodynamic relation $dG = VdP - SdT$, at constant temperature this equation can be rewritten as $(\partial \nabla G / \partial P)_T = \Delta V$, where $\Delta G$ is the increase of the Gibbs free energy in the volume $V_0$, and $\Delta V$ is the corresponding increase in volume, and $\Delta P$ represents the negative pressure (stress toward outward). Assuming that a high-energy electron with energy $2\Delta E$ traps itself in a confined volume $V_0$, then the volume change $\Delta V$ can be approximately written as

$$\Delta V = (\partial \Delta G / \partial V)_T (\partial V / \partial P)_T, \quad \text{hence}$$

$$\Delta V = [2 \Delta E (K \times V_0)]^{1/2} \qquad (4)$$

Where $K = -(\partial V / \partial P)_T / V_0$ is the compressibility. The increase in volume must lead to an increase of the internal energy, which should be equal to the work performed by the tensile stress from $V_0$ to $V'$, that is, $W = \int_{V_0}^{V'} P \, dV$ here $P = -(V - V_0) / K \times V_0$ (from Ref. 14). Thus the increase of the internal energy in the volume is

$$W = \Delta V^2 / (2K \times V_0) \qquad (5)$$

By comparing the equation (5) with the equation (4), it can be found that at the equilibrium condition, the increase in internal energy (strain energy) in the volume is $\Delta E$. This simple relation means that if the total transition energy of a high-energy nonbonding electron is $3\Delta E$, then the electron will use $\Delta E$ to create lattice strain, and $2\Delta E$ as free energy to maintain the lattice to stay at the new equilibrium position. Then the negative pressure occurring in the volume, $V_0$, can be obtained by the thermodynamic equation,

$$\Delta P = - (dW/dV)_T, \qquad \text{hence}$$

$$\Delta P = (2\Delta E / K \times V_0)^{1/2} \qquad (6)$$

Thus, the negative pressure acting on the confined volume is proportional to the square root of the free energy density induced by the trapped electron. In fact, the dynamic process of the high-energy nonbonding electron through lowering its free energy to trap itself in a potential well is a rather complicated self-consistent process which at present one has no available method to deal with. Our purpose here is just through some simple equations to qualitatively demonstrate that CIDSE does play a key important role in understanding the physical mechanism of superconductivity.

Figure 1(a) shows a typical energy structure required by superconductivity at the energies above the Fermi level in a given superconducting material. The high - lying level $E_i$ represents the energy state of a nonbonding electron before transition. The closely spaced levels overlapping with the three-dimensional potential well shown in Figure 1(a) are arisen from a widely dispersive antibonding band of the material. The bold lines under the antibonding band represent the three- dimensional potential well, which has a typical potential height of about 2 eV for cuprates. Here we need to note that the electronic states in the antibonding band are the unstably extended states, the electrons that only stay in antibonding bands are capable of carrying ballistic current in metallic and superconducting states. When there is existence of antibonding states at the energies above the potential wells as the case shown in Fig. 1 (a), at the temperature above $T_c$, the nonbonding high - energy electrons will first make a transition from $E_i$ to the antibonding states at the top of potential wells and becomes a conducting electron at the conduction band lying in energy far above the Fermi level. This is what we defined for the normal state of superconductors. Obviously, all of the transition energy of a nonbonding electron must be transferred into the lattice - strain energy and the kinetic energy of the transition electron (hot electron), since the energy carried by the nonbonding electron is a portion of the total intrinsic energy of the material. This transition energy for cuprates falls in a range from 1.5 eV for La214 to 8 eV for Bi-based compounds, estimated from the photoemission data in Ref. 2, 29. Thus we can see that the current in the normal state of the high-$T_c$ superconductors is conducted by the high-energy hot electrons at the energy levels far above the Fermi level of the material.

We need to keep in mind that the kinetic energies carried by this sort of conducting electrons cannot be transferred into thermal energy through interacting with phonons (harmonic vibrations). To achieve this requirement, the current in both normal and superconducting states of a given superconductor must be conducted by the electrons at the energy states far from the

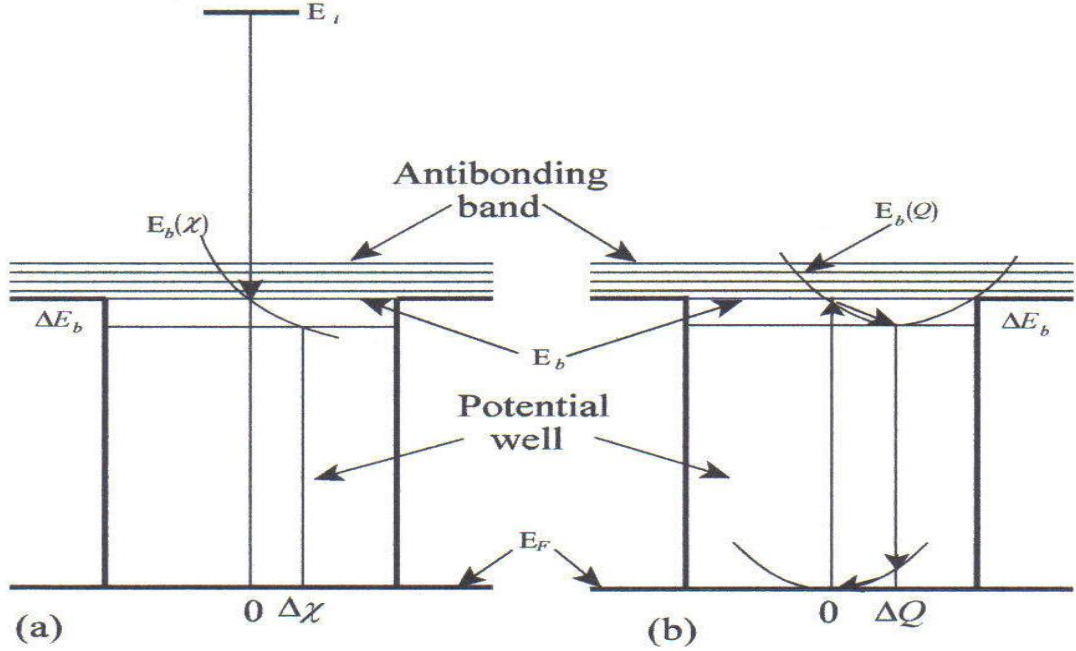

FIG. 1. The typical energy structure of superconducting materials. The diagram (a) shows schematically the three key elements for achieving superconductivity: the three-dimensional potential well located at above Fermi level, an antibonding band running over the top of potential well, and a high - lying level $E_i$ representing the energy of a superconducting electron before transition. The energy curve $E_b(\chi)$ expresses the changes of level $E_b$ with $\chi$, a parameter describing the change in volume of a potential well into which a superconducting electron is trapped. The energy shift $\Delta E_b$ denotes the binding energy of a superconducting electron in potential well. In diagram (b), the total energy, $E_b(Q)$, of an excited electron is expressed as a function of normal co-ordinate, Q. If a normal electron is excited from the Fermi level to the level $E_b$ by an external energy source, then the excited electron will relax to the state $E_b$ ($\Delta Q$) along the curve $E_b(Q)$ through emitting phonons, i.e. the Franck - Condon process. While in the former case (a), the energy shift $\Delta E_b$ is totally stored in lattice as an additional elastic energy.

Fermi level, or equally, far from phonon scatterings. The only energy that can be transferred from a conducting electron to the lattice is the thermal energy of an amount kT that the electron gained from the temperature field. Assume that J is the density of conducting current, and $\rho$ is the resistivity of the material, and then the power dissipated by the current in the unit volume is about $\rho J^2$, which should be equal to nkT, where n is the density of conducting electrons in normal state. If the current density J is maintained constant at various temperatures, then the resistivity of high-$T_c$ cuprates at the temperature above the transition temperature ($T > T_c$) will show a linear dependence on temperature, which is just what one has observed in all cuprate compounds with optimal compositions. Since the height of potential wells for trapping superconducting electrons in cuprates is too high to thermally excite the normal electrons from the Fermi level to the top of potential wells, so this linear behavior of the resistivity with temperature can last to more than 1000 K in cuprates with optimal doping.[15] However, the height of potential wells for trapping superconducting electrons is much lower in conventional superconductors (usually less than 0.1 eV), so with increasing temperature, a portion of the normal electrons at the Fermi level will be thermally excited into the energy levels at the top of the potential wells and conduct current with superconducting electrons together. For this reason, the linear dependence of resistivity with temperature usually cannot be observed at high temperatures in most of conventional superconductors. Thus we can expect that the higher the height of potential wells for trapping superconducting electrons, the larger the range for the linear dependence of the resistivity with temperature.

However, when the temperature is lowered below the transition temperature ($T < T_c$), it will be energetically favorable for the high - energy conducting electrons at free electron states through dynamic interaction with lattice to trap themselves into the underlying three-dimensional potential wells. Here we need to point out that by definition, almost all of the elementary excitations in solid state physics, such as polarons, bipolarons, polaritons, and self - trapped electrons in insulators etc. are all resulting from the electrostatic interaction between normal electrons and crystal lattice. The energy states for all these kinds of elementary excitations must be located at somewhere below the Fermi level. If the bound state is formed through the electrostatic interaction between electron and crystal lattice, then the bound state can be formed in any dimensionality (from d = 1 to 3).[16] However, the dynamic bound state which is formed by the high - energy nonbonding electron through dynamic interaction with its surrounding lattice can only be achieved in a three-dimensional potential well, and has an energy state at above the Fermi level. Just like in the case of a gas system, a thermodynamic equilibrium state can be only achieved in a three dimensional confined volume.

Suppose that the highest bound level $E_b$ of a potential well, which is arisen from one of the antibonding levels, has an energy as high as that of the height of potential well, and one high - energy nonbonding electron is trapped into it, then the trapped electron will use one third of its transition energy, $E_t = (E_i - E_b)$, to make the lattice of potential well an extended strain, due to the increase of free energy density in the confined volume, and the rest two third of the transition energy will be used to produce a negative pressure to maintain the expanded lattice at dynamic equilibrium position. After the electron traps itself in a potential well, the electron will undergo a strong interaction with the lattice. Then, it follows that the electron at level $E_b$ undergoes a red shift $\Delta E_b$ [see Fig.1(a)], as a result, the electron will become a dynamic bound one in the potential well.

In a system with strong electron-lattice interaction, a conventional method used to deal with electron-lattice interaction in solid state physics is to express the total energy of an electron as a function of the dimensionless normal coordinates, and the energy shift of a bound electron at the level $E_b$ caused by electron- lattice interaction is usually referred to as Franck-Condon shift, which can be approximately expressed as $E_{FC} = \Delta E_b = \sum_i S_i \hbar \omega_i$, where $S_i$ is the Huang - Rhys electron - phonon coupling strength for the phonon with energy $\hbar \omega_i$, and the sum is over all modes involved in the coupling.[10,17] This process means that a bound electron at level $E_b$ will relax to the stable state $E_b(\Delta Q)$ through emitting a number of phonons[18] as shown in Fig. 1(b). Clearly, this conventional method cannot be applied to the superconducting electrons, since superconducting electrons must prevent the interaction with any harmonic mode or phonon from the loss of their energy. That is, the superconducting electrons can only interact with lattice distortion or lattice strain to remain the total energy of a given material constant. In contrast, if an electron is excited from the Fermi level to the level $E_b$ by an external energy source, then this excited electron must follow the Franck-Condon process to relax to the final state by emitting phonons, as is the case we have observed in semiconductor nanocrystal systems.[10]

In condensed - matter physics, the energy shifts of the antibonding level $E_b$ has been represented as $\Delta E_b = D_c \times \Delta$, where $D_c$ is the deformation potential constant for a conduction band, and $\Delta$ is the dilatation in volume[13]. Here we use a single parameter $\chi$ to characterize the changes in volume of a potential well, and then the energy shift of a bound electron in a potential well could be expressed as a function of the parameter $\chi$ as shown in Fig.1(a). The configuration diagram on the base of the parameter $\chi$ shows that the energy shifts of a bound electron from the initial level $E_b$ to the final state $E_b(\Delta\chi)$ is totally stored in the lattice as an elastic energy. In the following section, we will see that it is just this elastic energy that can be periodically transferred between the superconducting electron and chain lattice to keep the supercurrent persisting.

According to the basic concepts of the electronic structure in solids, the key important feature of the antibonding states for achieving superconductivity is that the energy level for an antibonding state generally shows a down shift with increasing bond length. Thus an increase in the volume of a potential well will cause the antibonding level $E_b$ to have a down shift, which in turn ensures the electron transiting to the level $E_b$ from a high - lying level $E_i$ to become a dynamic bound state in the corresponding three - dimensional potential well. We refer to this dynamic bound state of superconducting electrons in the three-dimensional potential wells lying at above the Fermi level as superconducting state. The down shift of the level $E_b$, $\Delta E_b$, is defined as the binding energy of superconducting electrons, which dominates the critical transition temperature and the condensation energy of superconducting state in a given superconductor. Here it is worth noting that if level $E_b$ at the top of a potential well is resulting from bonding states, then the dynamic bound state cannot be formed, because the energy of an electron falling at a bonding state would uniquely shift to high energies with increasing bond length.

In principle, the energy shift $\Delta E_b$ can be calculated by using self-consistent method. However, the important task to us at present is through some measurable parameters to roughly estimate the magnitude of the energy shift $\Delta E_b$. For this purpose, the energy shifts of an electron at the antibonding level $E_b$ can be expressed as

$$\Delta E_b = (dE_b/dP) \times \Delta P, \qquad (7)$$

Where $dE_b/dP$ represents the pressure coefficient of level $E_b$ measured under an applied pressure. $\Delta P = -(2/3 E_t /KV)^{1/2}$ is the negative pressure induced by the trapped electron in the three-dimensional potential well. This simple equation clearly shows that the binding energy $\Delta E_b$ of superconducting electron will increase with increasing the transition energy of the nonbonding electron, $E_t$, and decrease with increasing the confined volume of a potential well.

In addition, the pressure coefficient of the superconducting state also plays an important role in determining the binding energy of superconducting electrons. It has been experimentally proved that the largest pressure coefficient for conduction band usually occurs at the highest symmetry point $\Gamma$ (s - symmetry) of the Brillouin zone, while the pressure coefficient at X point (p – symmetry) usually has a negative value[10]. Thus, the transition temperatures in p symmetry superconductors must be very low, even they do exist. Perhaps this is the reason why most of the conventional superconductors with high $T_c$ always show a superconducting wave with s - symmetry.

If we assume that the initial energy level $E_i$ of a nonbonding electron lie 1.5 eV above the potential well, and take $CuO_6$ octahedron as a potential well with a volume of $2.56 \times 10^{-23}$ cm$^3$, as is the case in La214, and suppose that a typical compressibility $K = 1.1 \times 10^{-2}$ GPa$^{-1}$ for cuprates, then the negative pressure induced by the trapped nondonding electron in the confined volume of $CuO_6$ is about 24 GPa, which in turn will lead to a dynamic bound state with a binding energy about 12 meV. During the estimation we already make an assumption that the pressure coefficient, $dE_b/dp$, of the antibonding level of Cu $d_{x^2-y^2}$ - O p$\sigma$ hybridized orbitals (pd$\sigma$) is about 0.5 meV GPa$^{-1}$, which is obtained by matching with most of the cuprates. If the potential well is changed from $CuO_6$ octahedron to $CuO_5$ pyramid, and the other parameters are remained the same, then the negative pressure induced by the trapped electron will be 33.9 GPa, which will lead to a superconducting state with a binding energy as large as 17 meV. The binding energy of the superconducting electrons in fact equals the thermal excitation energy of a superconducting electron from its bound state to the lowest free electron state. The thermal excitation probability of the superconducting electron is proportional to the Boltzmann factor, $\exp(-\Delta E_b / kT)$. At present, it seems impossible to predict $T_c$ accurately for all materials by a unique equation. For this reason, we may continue to use the empirical relation of the BCS model and assume $\Delta E_b = 3.5 \, kT_c$, where we use the binding energy of superconducting electrons to replace the energy gap of the BCS model. In this case, the binding energy of 12 meV and 17 meV will give rise to a transition temperature, 41 K and 58 K in the corresponding materials respectively.

According to the types of potential wells, we may refer to the conventional superconductors as those in which the potential wells are formed by the microstructures of materials, such as crystal grains, clusters, nanocrystals, superlattice, and the charge inversion layer in metal surfaces. The common feature for this sort of potential wells is that the volume of potential wells for confining superconducting electrons is varied with the techniques using to synthesize the materials. For this reason, the superconducting transition temperature in most of conventional superconducting materials is strongly sample - dependent and irreproducible.[19] The high-$T_c$ conventional superconductors mostly lie in the amorphous thin films synthesized by evaporating metals on cooling substrates, or the metal films sputtered by noble gases, or implanted by nitrogen ions. All these techniques are favorable to make the thin films with small crystal grains. It has been found that the smaller the size of grains or clusters, the higher the transition temperature of the material. In the following section, we will see that the electron-doped cuprates and $BaK(Pb,Bi)O_3$ compound, which have been accepted as high - $T_c$ superconductors, in fact, both belong to conventional superconductors. In contrast, if the potential wells for trapping superconducting electrons are uniquely formed by lattice structure of the materials, such as $CuO_6$ octahedrons and $CuO_5$ pyramids for cuprates, $BiO_6$ octahedrons for $(BaK)BiO_3$, and $FeAs_4$ tetrahedra in LaOFFeAs, or $C_{60}$ in $A_3C_{60}$ fullerides, no matter what the transition temperature is, we refer to this sort of material as high-$T_c$ superconductors in which the transition temperature is completely dominated by the intrinsic feature of the material lattice structure.

It has been long known that the high-pressure experiments can provide important information for searching superconductors with even higher $T_c$. The discovery of $YBa_2Cu_3O_7$ compound with $T_c$ above 90 K is a good example. From the simple equation (7) for determining the binding energy of superconducting electrons, we can see that an applied pressure can change the transition temperature of a superconductor through the following processes. Under an applied pressure, the increment in volume of the potential well caused by the trapped superconducting electron in a given material will be suppressed, which in turn will lead to a decrease in the binding energy of superconducting electrons. We have seen from the examples above that the negative pressure induced by a trapped electron in the confined volume is higher than 24 GPa for most of cuprates. Thus, at low pressure range, the effect of an applied pressure on the volume of potential wells is less important. However, the applied pressure can make the rest volume of the unit cell except the confined volume have a relatively large decrease, which may greatly increase the energy of nonbonding electrons, since the nonbonding electrons usually come from the antibonding states of metal ions. If this is the case, then the even higher value of $T_c$ may be achieved for a given superconductor by increasing the energy of the nonbonding electron through chemical substitution. However, the situation is quite different for conventional superconductors; the conventional superconductors generally have a large confined volume for superconducting electrons, which is close to the cube of one half of coherence length for most of conventional superconductors. The coherence length for conventional superconductors normally lies in a range from 10 nm to $10^3$ nm. Thus the corresponding volumes for trapping superconducting electrons fall in a range from $10^{-18}$ to $10^{-12}$ cm$^{-3}$ which is about five to ten orders of magnitude larger than that of the high-$T_c$ cuprates. Suppose that the energy of the nonbonding electron has the same value in both high-$T_c$ cuprates and conventional superconductors, then the negative pressure induced by superconducting electron in conventional superconductors is about two to five orders of magnitude smaller compared to that occurred in the cuprates. Thus, the confined volumes of superconducting electrons in conventional superconductors can be easily compressed by an applied pressure. So it is not surprising that the transition temperature for most of conventional superconductors shows a negative pressure derivative.

The alkali - doped $C_{60}$ compound is a completely new class of superconductor with a number of special properties, which provide another interesting example for further testing the CIDSE model. When the $C_{60}$ molecules condense into a solid, the weak interaction between the $C_{60}$ molecules will develop the discrete levels of a free $C_{60}$ molecule into the narrow bands with a typical width of about 0.5 eV. An undoped $C_{60}$ solid is a band insulator with a band gap about 1.6 eV, measured by photoabsorption. It is widely accepted that as solid $C_{60}$ is doped by alkali atoms, the alkali atoms donate one electron each to the conduction band, so the doped $C_{60}$ solid should be metallic.[20] However, it has been found experimentally that all alkali fullerides $A_n C_{60}$ (A is an alkali atom: K, Rb or Cs, and n is an integer from 1 to 6), only $A_3 C_{60}$ fullerides show the metallic behavior, while the others all behave as an insulator. It was also identified that for all electron - doping $C_{60}$ compounds, the superconductivity can only be observed in the range of 2.5 to 3.5 electrons per $C_{60}$ molecule.[21] In addition, the $A_3 C_{60}$ compounds also show some other puzzling features, particularly like that the superconducting transition temperature rises with increasing lattice constant, and the intramolecular vibration modes play the most important role in the electron - phonon interaction in their superconducting state.

Obviously, a right superconductivity model must be able to answer the fundamental question why superconductivity in the electron - doping $A_n C_{60}$ compounds only appears at the range around three electrons per $C_{60}$ molecule, and also can consistently explain the other puzzling features together. It is straightforward that in the $C_{60}$ compounds, every $C_{60}$ molecule itself is a potential well. If a $C_{60}$ solid becomes metallic, the conducting electrons in the compound should have energy higher than the height of $C_{60}$ potential wells. In other words, the value of the height of $C_{60}$ potential well should equal that of the energy gap of $C_{60}$ compound. That is, the bottom of $C_{60}$ potential well is located at the Fermi level of $C_{60}$ compound, just like the case shown in Fig.1(a). If we use the one - dimensional infinite potential well to roughly estimate the energy structure occurred in the real $C_{60}$ potential well and take the width of the infinite potential well equal to the diameter of a $C_{60}$ molecule sphere (0.71 nm), then the allowed energy states for n = 1 and n = 2 in a $C_{60}$ potential well have the value of 0.75 eV and 3 eV, counted from the Fermi level, respectively. Although the estimated allowed energy states above may much differ from those occurred in the real $C_{60}$ potential well, this simple estimate does provide an insight into the physical picture happened in $C_{60}$ compounds. The s - electrons of the alkali atoms in $AC_{60}$ and $A_2C_{60}$ compounds should first fill the ground state (n = 1) of $C_{60}$ potential wells, which can accommodate two electrons with opposite spin. Since the n = 1 ground state is deeply confined in $C_{60}$ potential well, or in other words, is located in the band gap, so it is not surprising that both compounds above behave like an insulator. However, in $A_3C_{60}$ compounds, one s - electron in each $C_{60}$ molecule is expected to occupy the levels at the top of $C_{60}$ potential wells, the conduction band, so the normal state of $A_3C_{60}$ compounds shows metallic behavior. As the temperature is lowered below $T_c$, the high - energy s - electrons in the conduction band will transit to the level $E_b$ and becomes a dynamic bound electron in $C_{60}$ potential well, just like the process as shown in Fig. 1(a). From this convincible example we can see that if there exist the low - lying stable states in potential wells at the energies below level $E_b$ in a given material, like in the case of crystal grain and nanocrystal potential wells, the high - energy nonbonding electrons have to fill them first, then the remains have a chance to become a dynamic bound state at level $E_b$.

Based on the CIDSE model above, it is obvious that the main reason for having a lattice expanding (shrinking) with doping in any material must lie in the increase (decrease) of free energy density with doping. It is well known that the ionization energy of an atom is defined as the minimum energy required to ionize the atom. So it is naturally follows that the smaller the ionization energy of the doping atom, the larger the contribution of the valence electrons of the doping atom to the free energy density of the doped material. Since the ionization energy for K, Rb and Cs is 4.3, 4.1, and 3.9 eV, respectively, so in $A_3C_{60}$ compounds, both the free energy density and the binding energy of s - electron in $C_{60}$ potential wells all increase with the size of alkali atoms. Thus, the final result leads to the behavior that the superconducting transition temperature in $A_3C_{60}$ increases with expanding lattices. Now, we can understand why the transition temperature for $A_3C_{60}$ fullerides all has a large negative pressure derivative, and why changing lattice constant by physical and chemical pressure cannot lead to the identical results for a given $A_3C_{60}$ compound.[22]

Since the superconducting state of $A_3C_{60}$ compounds is formed by the dynamic bound state of the high - energy s - electrons of alkali atoms in $C_{60}$ potential wells, so the electron - lattice interaction in $A_3C_{60}$ superconductors must be dominated by the intermolecular modes. Based on the superconductivity mechanism discussed in the following section, we can see that $A_3C_{60}$ belongs to the regime of high - $T_c$ superconductors, and has a coherence length $\xi = 2a$, here a is the lattice constant of $A_3C_{60}$ compounds. Now we have to point out that according to the superconductivity model we propose in this paper, the optimal electron concentration in $A_3C_{60}$ compound should be around n = 2.5, not at 3, and when n ≥ 3, the superconductivity must be completely suppressed. We propose that this discrepancy may be caused by the reason that a part of alkali atoms in $A_3C_{60}$ compounds perhaps is not completely ionic, especially for those located in the octahedral holes, since the radius of an octahedral hole (2.1 Å) is a little larger than the ionic radii of alkali ions, which are 1.33, 1.48 and 1.67 Å for K, Rb and Cs, respectively[22].

It is important to note that during the derivation of the equation for the negative pressure induced by superconducting electrons, it was already implied that the trapped electron distributes its energy uniformly in the entire volume of a potential well. In fact, it has been experimentally found that the type of the expanded strain caused by a dynamic bound state generally depends upon the symmetry of the electron state into which the nonbonding high - energy electron is trapped.[10] For instance, the excited electron in the Γ valley (s symmetry) of conduction band in GaAs nanocrystals produces a hydraulic s-symmetry strain. While an excited electron in an X valley (p symmetry) of conduction band in silicon nanocrystals will create a p-symmetry uniaxial strain.[10] Thus we can make a convincing assumption that the symmetry of the dynamic bound state dominates the type of the superconducting wave in the corresponding material. So it follows that most of conventional superconductors, such as simple metals, transition metals, the compounds based on transition metals and compounds with A-15 structure should show s-symmetry superconducting waves. The high-$T_c$ superconductor $BaKBiO_3$ and LaOFFeAs should also display an s-wave, while the cuprate compounds all exhibit a superconducting wave with $d_{x^2-y^2}$ symmetry.

Another important assumption which need to be addressed is that if there are no antibonding states at energies above the potential wells, or the highest antibonding state of the conduction band in a material is confined in the potential wells, then the energetically stable states for the nonbonding high-energy electrons perhaps are to transit to the low-lying available bonding states below the Fermi level, and all of the transition energy will be transferred into the localized bond energy, which in turn will lead to the stable lattice distortion or phase transition. This phenomenon is just what one has observed in the phase diagram of all cuprate compounds.

Since the energy carried by the high-energy nonbonding electron in solids are totally ignored in condensed-matter physics, so the phenomena arising from this kind of energy all become the big puzzles from the point of view of the theories standing on Landau - Fermi liquid theory. However, if we consider the energy carried by the high-energy nonbonding electrons as a portion of the total energy of a solid, and take superconducting state of a material as the dynamic bound state of the superconducting electrons, then almost all of the remarkable features of the superconductivity can be consistently explained by CIDSE model. In the following section we will illustrate the physical mechanism of superconductivity in terms of two typical high-$T_c$ superconductor $La_{2-x}M_xCuO_4$ and $Ba_{1-x}K_xBiO_3$. These two compounds have completely different physical properties, but under hole- doping, they both satisfy the three necessary conditions required by superconductivity and have the same superconducting mechanism.

### III. THE PHYSICAL MECHANISM OF SUPERCONDUCTIVITY

A fact which has long been known is that the superconducting properties of the materials are essentially governed by the electronic and lattice structure in their normal state. Here we will first briefly present the crystal and electronic structure on both $La_{2-x}M_xCuO_4$ (M= Ba, Sr and Ca) and $Ba_{2-x}K_xBiO_3$ compounds. We believe that the real mechanism for superconductivity must be hidden behind the common features of these two quite different compounds.

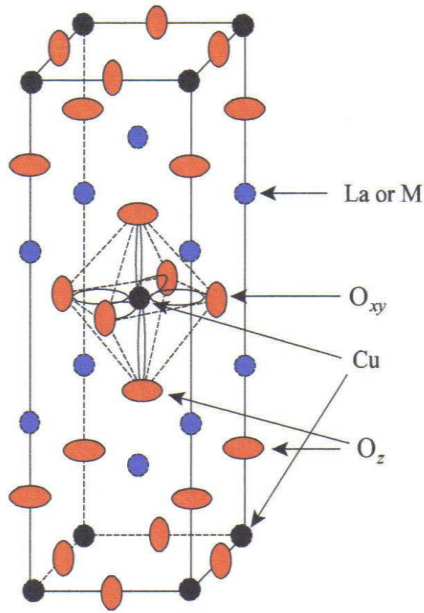

FIG. 2. (Color online) The crystal structure of $La_{2-x}M_xCuO_4$ compound in tetragonal phase. The ellipsoids show both dynamic and static off - site displacements of oxygen ions from their ideal sites. The vital structure for superconductivity in this compound lies in $CuO_6$ octahedron potential wells.

The tetragonal structure of $La_2CuO_4$ (La214) shown in Fig.2 has two nonequivalent O sites, in which $O_{xy}$ labels the O sites located in the $CuO_2$ planes and have a short Cu-O bond length (1.9Å), and $O_z$ denotes the apical O atoms with a Cu-O bond length (2.4 Å). When the temperature is lowered blow 500º K, the La214 makes a phase transition from tetragonal to orthorhombic symmetry. The distortion accompanied by the phase transition consists primarily of a rigid rotation of octahedrons by 5º around the tetragonal axes [110] or [1-1 0]. The pure La214 has an antiferomagnetic insulator behavior. It is widely accepted that the large differences in the bond lengths cause the compound to show a strongly anisotropic in its electrical properties. The superconductivity can only occur in the $CuO_2$ planes. When the doping x in $La_{2-x}M_xCuO_4$ is more than 0.05, the compound becomes metallic in $CuO_2$ planes and stabilizes the tetragonal structure at room temperature. It was found that in the tetragonal lattice structure of $La_{1.85}Ba_{0.15}CuO_4$, the bond length of Cu-O bonds in a square planar coordinated decreases from 1.9 Å of the pure La214 to 1.8936 Å, and the bond length for the apical Cu-O bonds increases to 2.43Å from 2.4 Å in La214. That is, the $CuO_6$ octahedrons in the hole-doped compound $La_{2-x}M_xCuO_4$ become more elongate than that in the pure La214. The Superconducting transition temperature depends critically on the concentration of the divalent atoms, and the general tendency of the $T_c$ with x in these compounds is quite similar for all divalent doping elements. This fact means that the important effect of doping on the superconducting properties of the La214 based compound essentially does not lie in the nature of doping divalent elements, but in removing a 5d electron of atom La from the corresponding unit cell. Another important feature which has been found in both pure and divalent atom doped La214 compounds is the unusually large and nearly temperature- independent ellipsoids occurred on the O ions in $CuO_6$ octahedrons, which mean that Cu ion is located in an octahedral potential well formed by the O anion ions.

The discovery of superconductivity in $Ba_{1-x}K_xBiO_3$ compound plays an important role in finding the unified mechanism of superconductivity, because this compound has neither copper nor antiferromagentism phase. This fact means that if the superconducting mechanism is the same in both copper and non - copper oxide compounds, then magnetism could not become a dominant issue in the unified mechanism of superconductivity. The pure $BaBiO_3$ has the lattice structure of cubic perovskite. The doped compound $Ba_{1-x}K_xBiO_3$, like in the case of $La_{2-x}M_xCuO_4$, undergoes several phase transitions depending on the concentration x of doping ions. As is the case for LA214, the pure $BaBiO_3$ shows a semiconductive character with a gap of about 2 eV. At a potassium concentration x > 0.37, the doping compound becomes metallic, and has a superconducting transition temperature T = 30 K (Ref. 23). As in the case of $La_{2-x}M_xCuO_4$, the lattice constant of $Ba_{1-x}K_xBiO_3$, a , smoothly decreases with increasing doping concentration x, which follows the formula, a = 4.3548 - 0.1743x (Å) (Ref. 24). The ion radii of $K^+$ and $Ba^+$ are 1.64 Å and 1.61 Å, respectively. Therefore, we can see that the lattice shrink with doping concentration is clearly not caused by the ionic size of the doping atom, but by the decrease in free energy density due to the loss of a Ba 6s electron in the doped cell. It is also observed that the oxygen ions in the $BiO_6$ octahedrons show the large and nearly temperature-independent ellipsoids in the planes of cubic faces, as shown in Fig. 3. It follows from this fact that the oxygen ions form a cubic-symmetry potential well surrounding a Bi ion.

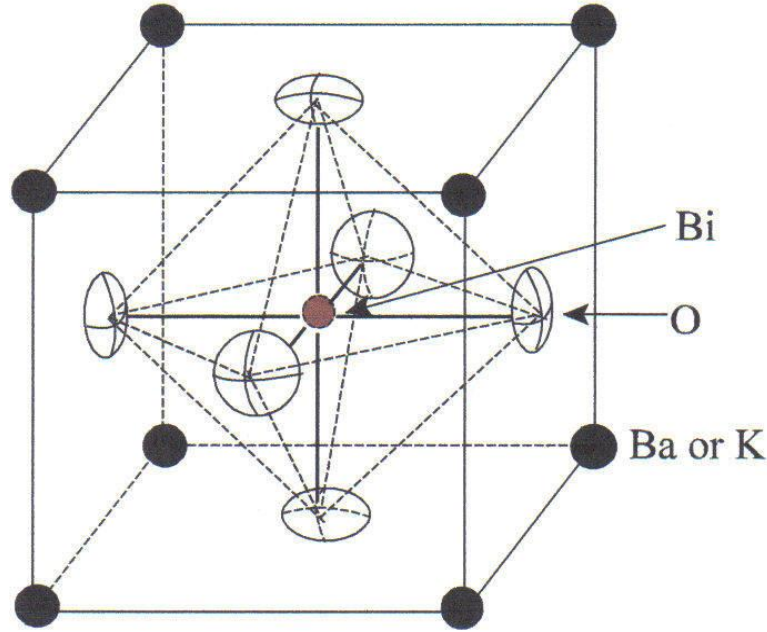

FIG. 3. (Color online) The lattice structure of (K, Ba)$BiO_3$ in the cubic phase.

The ellipsoids represent the displacements of oxygen ions from their ideal sites. The key structure for superconductivity in this material, like in La214 compound, also lies in the $BiO_6$ octahedron potential well.

As mentioned above, the normal-state properties of both compound $La_{2-x}M_xCuO_4$ and $Ba_{1-x}K_xBiO_3$ are similar in several aspects. It can be expected that the common features in both compounds must relate to the octahedral structure of oxygen ions surrounding a Cu or Bi ion, since the electronic band structure in both compounds are essentially dominated by their corresponding octahedral lattice structure.

The band structures for both compounds were first reported by Mattheiss.[25,26] In cubic crystal field, the five degenerate 3d states on Cu ion in La214 split into a doublet, $e_g$( $d_{x^2-y^2}$, $d_{3z^2-}$ $_r^2$) and a triplet $t_{2g}$( $d_{xz}$, $d_{xy}$, $d_{yz}$). As the symmetry of $CuO_6$ decreases from cubic to tetragonal, the 3d states will further split into $b_{1g}$ ($d_{x^2-y^2}$), $a_{1g}$ ($d_{3z^2-r^2}$), $b_{2g}$ ($d_{xy}$) and $e_g$( $d_{xz}$, $d_{yz}$). While the degenerate oxygen 2p states p(x), p(y) and p(z) will split in the crystal field of the $D_{2h}$ into three levels: (p$\pi_{//}$), (p$\pi\perp$), (p$\sigma$). The five Cu 3d states and the three O 2p states for each of the four oxygen atoms in plane form a complex of 17 hybridized p - d valence bands centered at - 3 eV below $E_F$. The particularly important band for superconductivity is the antibondig band resulted from Cu $d_{x^2-y^2}$ and O 2p$\sigma$ orbitals (pd$\sigma$ band), which is primarily formed by Cu $d_{x^2-y^2}$ orbitals.

This half filled antibonding band crosses $E_F$ and has a maximum at about 1.8 eV above $E_F$ located at X point of the Brillouin zone (estimated from the diagram of band - structure in Ref. 25). The La 5d states starting around 1 eV merge with the La 4f bands at 3 eV above $E_F$. The La and $O_z$ ions are fully ionic. That is, there is a full charge transferring from La atoms to $CuO_6$ octahedrons. An important feature to note is that the La214 compound contains odd number electrons in its unit cells, which means the antibonding band pd$\sigma$ crossing $E_F$ is half filled. On the basis of energy - band theory, this character ensures the pure La214 compound being metallic. However, the resistivity measurements have shown that the pure La214 compound is an insulator with an energy gap of about 2 eV. In contrary to the resistivity measurement, Fermi surfaces were observed by Tanigawa in terms of positron date on La214, which means that the La 214 has metallic character with special scattering centers.[27] Later, it was reported that the $La_2CuO_{4+y}$ also exhibits superconductivity near 40K. It has been identified that the extra oxygen occupies interstitial positions in the La-O layers.[23] In this case, the extra oxygen cannot make the band structure around the Fermi level in the doped La214 any different from that of the pure La214 compound. This fact demonstrates that the insulator - metal transitions in the La214 based compounds do not depend on their band structure around the Fermi level.

The remarkable superconductive phenomena observed in $Ba_{1-x}K_xBiO_3$ compound are qualitatively similar to those found in $La_{2-x}M_xCuO_4$ system. The essential feature of the band structure in the $BaBiO_3$ compound lies in the widely dispersive antibonding band derived from a strong hybridization of Bi 6s - O2p$\sigma$ orbitals. The Bi 6s - O 2p$\sigma$ antibonding band, primarily of Bi 6s orbitals, in pure $BaBiO_3$ compound has a maximum of about 1.7 eV above $E_F$ located at M point of Brillouin zone, estimated from the band structure.[26] Both the widely dispersive antibonding band crossing Fermi level and an odd number of electrons in the primitive cell determine that the normal state of $BaBiO_3$ should be metallic. However, the measurements of the resistivity on pure $BaBiO_3$ sample exhibits an insulator with an optical gap of about 2 eV. The insulated feature of the pure $BaBiO_3$ compound in monoclinic phase has been accounted for by the formation of the charge density wave in the non-equivalent Bi sites. While the insulated behavior is also found in orthorhombic phase of pure $BaBiO_3$, in which the non - equivalent Bi sites are not observed. Like in the case of La214, the $Ba_{1-x}K_xBiO_3$ compound also shows an insulator - metal transition depending on doping. However, according to the rigid band model, the replacement of Ba by K has no effect on the band states near Fermi level. Thus, we can draw the conclusion that the insulator - metal transition with hole doping in both compounds does nothing with the band structure in the vicinity of the Fermi level in both materials.

Currently, the insulated behavior observed in cuprates has been ascribed to the Mott insulator. The basic idea for Mott insulators is that each copper atom in cuprates has a loose bound valence electron which creates a strong on site repulsion potential to prevent the electrons from hopping between copper ions. So the cuprates as a whole shows an insulated character. While, when the divalent atoms replace some of La atoms, the divalent atoms have a greater affinity for electrons and so attract the loosely bound electrons from the copper ions. Due to the loss of loosely bound electrons on some copper sites, the loosely bound electrons have chance to hop between the copper ions and carry electricity[28]. The problem for this model is that the energy state for the loosely bound valence electron on a copper ion must be located below Fermi level. While the energy levels of the divalent ions should be close to that of La ions, which at least lay 1 eV above Fermi level. Thus, it seems energetically impossible for an electron at copper ion to automatically transfer to the near dipositive ions. In addition, the $Ba_{1-x}K_xBiO_3$ compound also exhibits an insulator-metal transition with hole doping. In this compound, there is neither copper nor other transition metal, so the Mott insulator model is by no means a common key to open these two mysterious doors together.

We propose that the discrepancy between the theoretical prediction and experimental measurements occurred on these two compounds is caused by $CuO_6$ or $BiO_6$ octahedral lattice structure in their corresponding compound. The large ellipsoids resulted from dynamical displacements of oxygen ions in $Cu(Bi)O_6$ octahedron construct a potential well in which the conducting electron in the corresponding antibonding state is strongly confined. If the height of $Cu(Bi)O_6$ potential well is higher than the maximum of the antibonding band in the corresponding material, then the compound will show an insulator behavior in its normal state. The height of the potential well formed by oxygen ions in an octahedral coordination can be roughly estimated by putting one electron charge uniformly on a shell surrounding a Cu ion with radius R = 2.4 Å. Then an electron inside the shell will feel a potential height as high as $e^2 / R = 5.8$ eV. This estimate has already neglected the effect of cation within the shell on the potential height. Obviously, the value obtained for the height of $CuO_6$ or $BiO_6$ potential well is a little higher than what one has observed in the real cases. But this simple estimate demonstrates that the potential wells formed by the octahedral lattice structure of oxygen ions do play an important role in the electronic properties of both compounds. In fact, the distribution of charge density on a real $CuO_6$ or $BiO_6$ potential well apparently is not uniform. The charge density at the areas of the ellipsoids of oxygen ions is definitely higher compared to the other areas, which must lead to a correspondingly large potential height in those areas. Especially in $O_z$ directions, the potential height may be even higher than that obtained by the simple estimate, due to the valence state of 2- in $O_z$ ions. Following this fact, the La214 based compound must show strong two-dimensional character as observed experimentally. Since the charge density at the areas between ellipsoids of oxygen ions should be relatively lower compared to that at the ellipsoid areas, so we can see that even in $CuO_2$ planes, superconducting properties may also vary with direction. Based on this new model, it is reasonable to take the energy gap of 2 eV measured by optical conductivity as the lowest height of $CuO_6$ potential well along $CuO_2$ plane direction. As noted above, the maximum of pdσ antibonding band lies about 1.8 eV above $E_F$ estimated from the band structure calculated by Mattheiss.[25] That is, the antibonding state, which is the only state that can conduct the ballistic current, is confined in $CuO_6$ potential wells, so it is not surprising that the normal state of La214 compound shows an insulator character.

The $Ba_{1-x}K_xBiO_3$ has a cubic symmetry in its superconducting phase, which means that the charge density on 6 ellipsoids of $BiO_6$ potential well must be the same, so the potential height in the cubic face directions should have the same value. But the potential height at between ellipsoids of oxygen ions has relatively lower value. Thus the superconducting properties, such as coherence length and binding energy, also vary with directions. Assuming that the optical gap of 2 eV obtained in pure $BaBiO_3$ compound is the lowest height of $BiO_6$ potential well, then it is apparent that the highest energy state of the Bi 6s - O 2pσ antibonding band, which lies 1.7 eV above $E_F$, is strongly confined in $BiO_6$ potential wells. Thereby, there is no doubt that the normal state of pure $BaBiO_3$ shows an insulator character.

It is important to keep in mind that the replacement of Ba by K, or hole doping, has two functions that can critically affect the superconducting properties. First, as noted above, the lattice constant which is twice as long as the Bi -O bond length, smoothly decreases with increasing concentration of potassium, which in turn push the antibonding state of Bi 6s - O2pσ orbitals, primarily of Bi 6s orbital, up in energy. Secondly, under hole doping, the charge density on the shell of a potential well dramatically decreases, which will greatly suppresses the potential height. It is clear from foregoing discussions that both functions of hole doping are favorable to the insulator - metal transition of $Ba_{1-x}K_xBiO_3$ compound. It has been noted that a superconductor must satisfy three necessary conditions. For $Ba_{1-x}K_xBiO_3$ compound, we already have potential well $BiO_6$ for trapping superconducting electron, and the high-energy electron in this system is provided by 4s electron of K atoms, which has an ionization energy of 4.3 eV, while 7.3 eV for Bi atom. As long as the antibonding state of Bi 6s - O2pσ orbitals runs over the lowest height of $BiO_6$ potential well, the superconducting state can be formed in $Ba_{1-x}K_xBiO_3$ compound. It has been found that metallic conductivity in $Ba_{1-x}K_xBiO_3$ does not occur until a potassium concentration $x = 0.37$. At this doping concentration, the maximum of the conduction band at the M point runs over 2 eV above the Fermi level, estimated from the band structure in Ref. 26. Since this doping concentration is already over the optimal one ($x = 0.3$) required by the minimum coherence length ($\xi = 2a$). So it is not surprising that the transition temperature in $Ba_{1-x}K_xBiO_3$ compound shows monotonic decreases with further increasing doping. In order to make the discussion more clear and convincing, in the remaining portion of this section, we will confine our mind mainly on La214 compound to illustrate the physical mechanism of superconductivity.

The atomic configuration of the element lanthanum is $[Xe](5d)(6s)^2$, in which the 6s electrons has an ionization energy of 5.6 eV. Notice that atomic d - states are confined more closely to the nucleus than is s state of the same energy, so that in forming $La_2CuO_4$ compound, the two 6s electrons at La atom will first hybridize with $O_z$ 2p electrons, which have an ionization energy of 13.5 eV. Consequently, the average electron energy at $O_z$ ions become much higher than that of oxygen atom $O_{xy}$ in the $CuO_2$ planes since the ionization energy of Cu atom (7.72 eV) is much higher than that of La atom. For this reason, the bond energy of $O_z$ - Cu (d $z^2$) bond should also become higher than that of Cu – $O_{xy}$ bonds in $CuO_2$ planes. Thus, the $CuO_6$ octahedrons must undergo an elongation along $O_z$ - Cu ($dz^2$) bond direction. If the temperature is high enough (say, above $500°C$) to keep the nonbonding La 5d electron staying in its atomic orbital, then the pure $La_2CuO_4$ should show a tetragonal phase. Since the hybridization of La 6s electrons and $O_z$ 2p electrons greatly increases the site energy of the $O_z$ ions, so which in turn makes $O_z$ ions manifest themselves as unusually large dynamic off - site displacements from their ideal sites.

Based on the band model discussed above, the band structure of La214 compound is essentially dominated by the electronic structure of $CuO_6$ octahedra. Under this band picture, the nonbonding electron at La 5d electronic state is simply treated as a normal valence electron like others in valence band, and so the energy released by the La 5d electron transition is totally ignored. Since the energy carried by the nonbonding electrons is a portion of the intrinsic total energy of a solid, so the transition energy released by the high–energy nonbonding electrons must be transferred into other kind of internal energy. Since there are no free electron states at energies above $CuO_6$ potential wells in pure La214 compound as noted above, at low temperature, the nonbonding 5d electrons at La atoms will make transition to the low-energy $O_{xy}$ p(x, y) bond orbitals in $CuO_2$ planes, and release the transition energy to the corresponding bonds. This large transition energy (about 5 to 6 eV) greatly enhances lattice instabilities and also gives rise to the large dynamic displacements of $O_{xy}$ atoms from their ideal sites.[29] There are two La (5d) electrons in each unit cell. They will respectively occupy the empty $p\pi_{//}$ and $p\pi\perp$ bond orbitals. When the La (5d) electrons occupy $p\pi_{//}$ bond orbitals, most of their transition energy would be used to create a corrugation of the $CuO_2$ planes along the long axis direction. While, if the La (5d) electrons occupy the $p\pi\perp$ bond orbitals, they will bring a large amount of energy into the direction perpendicular to $CuO_2$ square plane, which would make $CuO_6$ octahedrons a rigid rotation around their symmetry axes. Then the correlative rotation of octahedrons will lead to a phase transition of L214 from tetragonal to orthorhombic.

However, when a trivalent La atom is replaced by a divalent atom ( Ba, Sr, Ca ), or in other words, hole doping, the unit cell containing a divalent atom lacks a La 5d electron in its $p\pi\perp$ orbital compared to other undoped cells. Thus, this cell may remain in its tetragonal phase due to the loss of rotation energy for its phase transition. With increasing the concentration of divalent atoms (say, x = 0.02), the material as a whole has no enough energy to make phase transition. That is the reason why hole doping has the function to stabilize La214 based compounds in tetragonal phase. After the doped La214 compound remains in its tetragonal phase, it is perhaps energetically favorable for the electrons at $p\pi\perp$ orbital in the undoped cells to become stable bound state in $CuO_6$ potential wells after they transfer most of their transition energy to $O_{xy}$ ions. We suppose that this stable bound state perhaps is the origin of the famous mid-infrared band observed by optical conductivity in the underdoped La214 compound, which has a binding energy of about 0.5 eV (Ref. 30) as shown in Fig.4. After the stable bound state is formed in $CuO_6$ potential wells, the charge density at the shell of potential wells is greatly decreased, which in turn would remarkably reduce the potential height along $CuO_2$ plane. At the same time, the valence state of copper ions transfers from $Cu^{2+}$ to $Cu^{1+}$. Following this change, the long - range antiferromagnetic phase in pure La214 compound disappears.

Under hole doping, the free energy density in the unit cell containing a divalent atom becomes lower compared to that of surrounding undoped cells, due to the loss of a high - energy La (5d) electron. According to the CIDSE model, this cell must suffer a compression from surrounding crystal field, which consequently leads to a decrease of the bond length of $Cu - O_{xy}$ bonds in the doped cell. Following the lattice shrink, the antibonding state Cu $d_{x^2-y^2}$ in the doped cell will move up in energy. Another function of hole doping, which may play a more important role compared with the former, is to reduce charge density on the shell of $CuO_6$ potential well, and in turn suppresses the height of potential wells along the $CuO_2$ plane. Hence, at a certain concentration of divalent atoms (say, x = 0.05), these two functions of hole doping will make the antibonding state in the doped cells run over the potential height. In this case, all unit cells doped by divalent atoms will exhibit the metallic character. If the average distance between two doped cells is smaller than the free path of electrons in $CuO_2$ planes, then the material as a whole becomes a metal. It is important to note that if we say some cuprate compound is in metallic state, it does not mean every cell of this compound being metallic, but only the doped cells are in metallic state. This coexistence phenomenon of metal and insulator, or superconducting and non-superconducting regions has been observed in both high-$T_c$ cuprates and conventional superconductors.[31]

According to the energy band structure calculated by Mattheiss,[25] the La (5d) electron states starting around 1 eV extend to 3 eV above $E_F$. While the direct and inverse photo emission data on $La_{1.85}Sr_{0.15}CuO_4$ have shown that La (5d) electron states extend up to about 5 eV above $E_F$ (Ref. 30). It should be reasonable to assume that the initial state of La (5d) electron lays 3.5 eV above $E_F$. On the other hand, the height of $CuO_6$ potential well in $CuO_2$ plane direction quickly decreases with increasing the concentration of divalent atoms. Suppose that at x = 0.125, the height of CuO$_6$ potential well decreases from 2 eV for pure La214 to 1.7 eV in CuO$_2$ plane direction as shown in Fig. 4, then the transition energy of a La (5d) electron from its initial state to the highest level E$_b$ of CuO$_6$ potential well is about 1.8 eV.

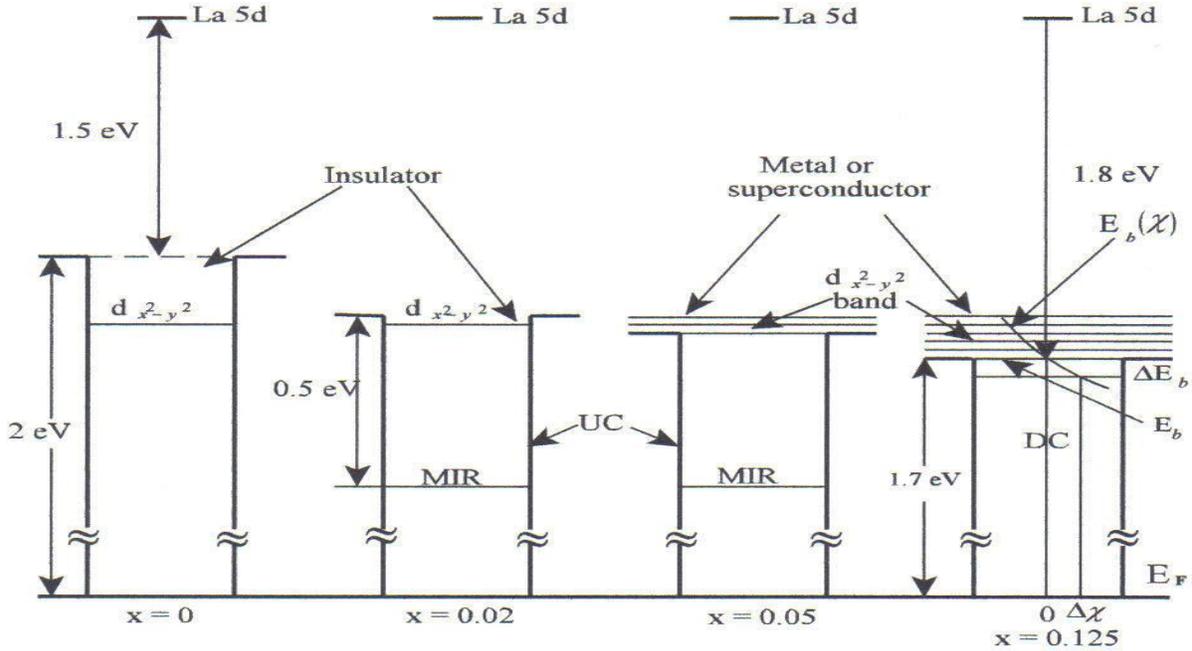

FIG. 4. The schematic diagram illustrating the origin of insulator-metal-superconductor transitions in La$_{2-x}$M$_x$CuO$_4$ compound with hole doping. Note as discussed in text that the potential wells shown in the diagram only represent those occurred in CuO$_2$ layers. The La 5d levels denote the initial energy of La 5d electron in the compound. $d_{x^2-y^2}$ represents the antibonding state of Cu $d_{x^2-y^2}$ - Op$\sigma$ hybridized orbitals (pd$\sigma$ band). When the antibonding state $d_{x^2-y^2}$ is confined in the potential wells, the material shows an insulator character.

While as the antibonting band runs over the height of the potential wells, the material will become a metal or a superconductor. MIR reflects the energy level responsible for the midinfrared band observed by the optical conductivity in La$_{2-x}$M$_x$CuO$_4$, UC - undoped cell, DC - doped cell.

Based on the CIDSE model, when a high-energy nonbonding electron traps itself in a potential well, the trapped electron has to distribute its energy along the symmetry of wave function of the bound state. Since the two - dimensional antibonding state Cu $d_{x^2-y^2}$ cannot stably confine a high-energy electron, we assume that the electron trapped in Cu $d_{x^2-y^2}$ antibonding state distributes its energy in the entire $CuO_6$ octahedron, which has a typical compressibility k = 1.1 x $10^{-2}$ / GPa. By the equation (6), the negative pressure induced by a La (5d) electron in a $CuO_6$ octahedron is about 26 GPa, which will lead to a binding energy as high as 13 meV for superconducting electrons. Here we need to note that this is only a rough estimate. In fact, most of the energy of a high-energy electron in Cu $d_{x^2-y^2}$ state is used to push the four nearest $O_{xy}$ ions in $CuO_2$ planes out - ward moving, which in turn must lead to an inward moving of the apical $O_z$ ions. Egami, et al have found by neutron scattering data that there exist displacements of atoms on $CuO_6$ octahedrons from their ideal sites and the Cu – $O_z$ separation is reduced to 2.16 Å in superconducting state of $La_{2-x}$( Ba, Sr$)_x CuO_4$ compound.[2,32]

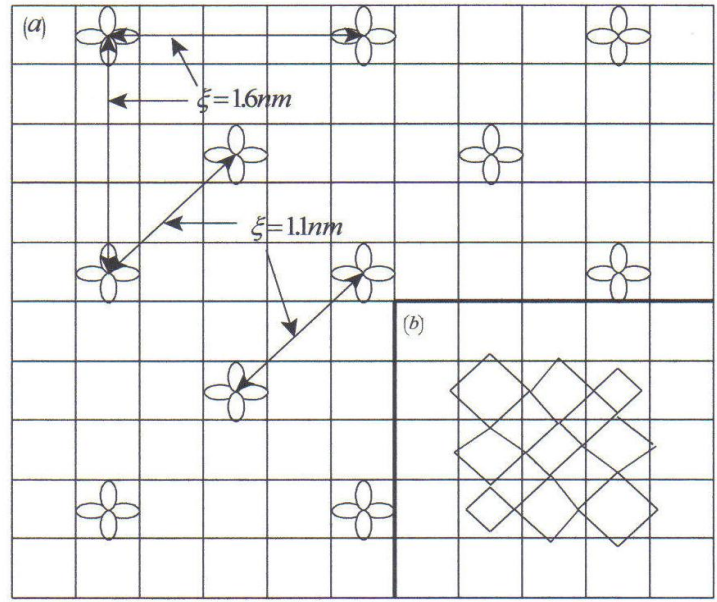

FIG. 5. The schematic diagram of the lattice structure of $CuO_2$ layers in $La_{2-x}M_xCuO_4$ (x = 0.125). The diagram (a) shows the normal state, and (b) indicates the possible lattice distortions occurred in superconducting state. The coherence length $\xi$ = 1.6 nm in [100] (or [010]) direction, and $\xi$ = 1.1 nm in [110] direction are shown in diagram (a), respectively. The unit cell with a Cu$d_{x^2-y^2}$ orbital in (a) indicates that the cell is doped by a divalent atom.

Now we arrive at the stage to illustrate the superconducting electronic process in terms of $La_{2-x}M_xCuO_4$ compound with x = 0.125. After $La_{2-x}M_xCuO_4$ compound becomes metallic and temperature is lowered below transition temperature, the remained La (5d) electron in the doped unit cells has a chance to become a dynamic bound one in the corresponding $CuO_6$ potential well. The lattice of Cu – $O_{xy}$ planes in normal state of $La_{2-x}M_xCuO_4$ is shown in Fig, 5(a), in which the Cu $d_{x^2-y^2}$ orbitals represent the unit cells containing a divalent atom. The possible distortions in $CuO_2$ plane caused by the trapped superconducting electrons in superconducting state are also shown in Fig. 5 (b). A superconducting chain in [100] or [010] direction is schematically shown in Fig. 6, which consists of a number of deformed $CuO_6$ octahedrons. The octahedrons which are expanded in their $CuO_2$ square plane in the chain represent that a superconducting electron is dynamically bound in each of them due to substitution of trivalent La by a divalent atom in the corresponding unit cells. The $CuO_6$ octahedrons lying between two expanded ones in the superconducting chain undergo a compressed deformation caused by the stress resulting from the expanded octahedrons in which the superconducting electron is trapped. The compressed deformation occurred in the middle one between two expanded octahedrons should be highest compared to others as shown in Fig.6 (a). Now, it is interesting to note that under a compressing stress aligned along the chain direction, a pdσ antibonding level will split into two, one arising from the Cu - O bonds aligned along the chain direction will move up with decreasing their bond lengths, while the other one resulting from the Cu- O bonds in perpendicular to the chain direction should have a down shift due to the elongation in their bond lengths. In this case, all unit cells in a superconducting chain become a metal along the chain direction, and an insulator in perpendicular to the chain direction. In other words, the electrons trapped in a superconducting chain can only move along the Cu - O bonds aligned in chain direction.

Now suppose that if a direct voltage is applied on the superconductor along the chain, then the dynamic bound electrons in $CuO_6$ potential wells will gain an energy from the applied voltage source and raise their energy state from the bound state $E_b(\Delta\chi)$ to a low - lying free electronic state as shown in Fig.4. Assume that the voltage dropped on the superconducting chain is $\Delta V$, then the energy gained by a bound superconducting electron is $\Delta V \times e$. In order to ensure a bound superconducting electron in $CuO_6$ potential well to become a conducting electron, it is necessary to make the energy, $\Delta V \times e$, a little higher than the binding energy of superconducting electrons. We refer to $\Delta V \times e$ as the excitation energy of the superconducting electrons. In fact, the excitation energy of a superconducting electron consists of both its binding energy and kinetic energy that the electron grained from its drift under an applied electric field. Once the trapped electron reaches a free electronic state under an applied electric field, it will become a conducting electron and begin to move into the nearest neighbor $CuO_6$ octahedron in the direction opposite to the electric field. At the same time, all the deformed $CuO_6$ octahedrons in the superconducting chain begin to restore toward their equilibrium position due to the loss of negative pressure created by the trapped electrons in $CuO_6$ octahedrons of the doped cells. Once the lattice equilibrium position is reached, the superconducting electrons should move into the octahedrons which are initially located at the middle. At time $t = T/2$, T is the oscillating period of superconducting electrons, all $CuO_6$ octahedrons occupied by superconducting electrons undergo an expanded deformation in which the four $Cu-O_{xy}$ bonds in $CuO_2$ plane are being elongated, while the two $Cu–O_z$ bonds will be constricted. At the same time, all superconducting electrons begin to release their excitation energy that they gained from external field to the chain lattice as an additional elastic energy. Consequently, all octahedrons between the two occupied by superconducting electrons suffer a compressed deformation as shown in Fig. 6(b). After superconducting electrons relax to their original bound level $E_b(\Delta\chi)$, the superconducting electrons at this time cannot stabilize at the bound state as in the case of the initial superconducting state, because the chain lattice gains an extra elastic energy from each superconducting electron. Thereby, the chain lattice will automatically return back to its equilibrium position, and following this process, superconducting electrons will take their excitation energy back from chain lattice (binding energy plus kinetic energy). Once the superconducting electrons get their energy back from chain lattices, they will move towards next octahedrons which are going to be expanded. At $t = T$, the system returns back to the case as the initial superconducting state as shown in Fig. 6(a). Then the system continuously repeats the compressing and restoring processes, even if the external electric field is removed. It is obvious that in order to ensure the superconducting process to persist in time, in the lattice restoring process, the energy gained by superconducting electrons from the chain lattice must be exactly equal to that they have released to the chain lattice during the compressing process. Clearly, the superconducting process is quite similar to the oscillation process of a mass on a spring with no loss in mechanical and thermal energy, the energy using to start the oscillation is periodically transferred between kinetic energy of mass and elastic energy of spring. While in superconducting process, the excitation energy of a superconducting electron is periodically transferred between free energy stored in superconducting electron and elastic energy stored in chain lattice. That is, the superconducting electrons transfer periodically between their dynamic bound state and conducting state.

Recently, the scientists at HZB have successfully found that in the superconducting process of $La_{1.2}Sr_{1.8}Mn_2O_7$ compound, the superconducting electrons periodically alternate between their free conductive state and trapped state.[33] The reversible process of the superconducting electron alternating between its trapped state and conductive state means that there must exist a fixed energy exchanged between superconducting electron and chain lattice. Obviously, the amount of this exchange energy should equal that of the excitation energy of superconducting electron in its dynamic bound state as defined above. Furthermore, the reversibility of a fixed energy transferred between superconducting electron and chain lattice also demonstrates that the superconducting electron – lattice interaction in superconductors should manifest itself neither electrostatic nor electromagnetic process, since the electron – lattice interaction energy for both cases above is stored in electromagnetic field, which in any case cannot be automatically transferred between superconducting electrons and chain lattice.

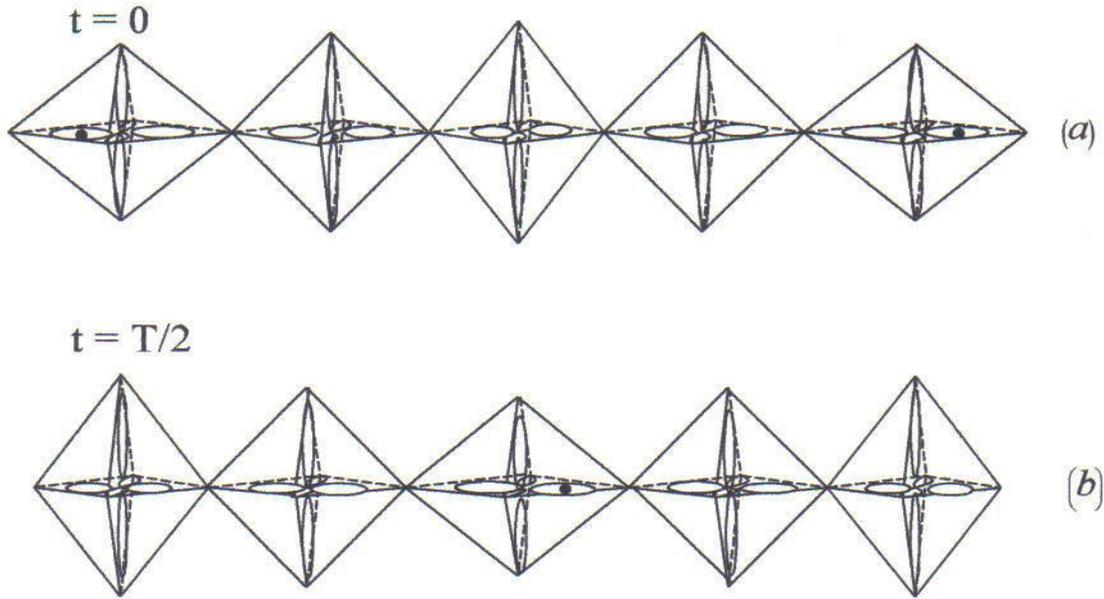

FIG. 6. A part of a superconducting chain along [100] or [010] direction in $La_{2-x}M_xCuO_4$ (x = 0.125). The diagram (a) and (b) show schematically the configuration of the superconducting chain at time t = 0 and t = T/2, respectively. The octahedron with a solid circle denotes a superconducting electron in its dynamic bound state. In order to keep the supercurrent persisting in time, the superconducting electrons must keep coherently moving with lattice distortion wave.

Thus, this important finding further confirms that superconductivity is undoubtedly caused by the dynamic interaction between superconducting electrons and chain lattice, which can only be achieved by the dynamic bound state formed by the high – energy nonbonding electron through dynamical interaction with its surrounding lattice to trap itself into an antibonding state located at the top of a three – dimensional potential well lying in energy at above the Fermi level. Now we can see that the intrinsic nature of superconductivity is to generate an oscillating current under a dc voltage. The wave length of oscillation current equals the distance between the two nearest bound superconducting electrons, namely the coherence length. In cuprate compounds, the coherence length just equals the distance between two nearest doped cells in the superconducting chain direction. The coherence length in $La_{2-x}M_xCuO_4$ compound is about 1.6 nm in [100] or [010] direction, and 1.1 nm in [110] direction for x = 0.125 as shown in Fig. 5.

It is straightforward from Fig, 5 and Fig. 6 that the coherence length in all cuprate compounds must satisfy the relation $\xi = 2na$ in $CuO_2$ planes, where n = 1, 2, 3, integer, a is lattice constant which is also equal to the width of potential wells if the electric field is applied along the axial directions of the unit cell. That is, the coherence length should have the value equal to an even number times the lattice constant. Thereby, in $La_{2-x}M_xCuO_4$ compound, the minimum coherence should be $\xi = 2a = 0.8$ nm, which corresponds to an optimal concentration of divalent atoms  $x = 0.25$.

The superconducting process with $\xi = 2a$ is the most common one, in which the superconducting chain is constructed by alternating arrangement of the expanded and compressed potential wells. Following the discussion above, when the superconducting electrons are excited from their dynamic bound state in the expanded potential wells into free electron states by an applied dc voltage, then the superconducting electrons in the doped cells are able to move into their nearest neighbor compressed one (undoped) at $t = T / 2$, then all potential wells in the chain begin to restore toward their equilibrium position and at the same time, the superconducting electrons begin to release their excitation energy to the chain lattice. After the superconducting electrons transfer all their excitation energy to the chain lattice, the chain lattice begins to restore toward their equilibrium position, and following this process, the superconducting electrons begin to take their excitation energy back from the chain lattice. When the chain lattice reaches to its equilibrium position, the superconducting electrons completely get their excitation energy back and start moving to the nearest potential well at $t = T$. Then the superconducting process above continuously repeats and the supercurrent persists without any dissipation. We propose that all conventional superconductors work in the superconducting process with $\xi = 2a$, here a is the width of potential wells.

Clearly, the maximum value of supercurrent density in cuprates should be also occurred at the doping concentration $x = 0.25$. Over this concentration, both the density of supercurrent and the transition temperature will be sharply suppressed by further increasing doping, since the extra divalent atoms will destroy the superconducting processes surrounding them. That is the reason why the superconductivity in all cuprate compounds completely ceases as hole - doping concentration $x > 0.3$. While the longest coherence length occurred in axial directions should close to $\xi = 4a = 1.6$ nm, which corresponds to $x = 0.0625$. Based on the CIDSE model discussed above, in the underdoped region, the height of $CuO_6$ potential well has a relatively large value, which will give rise to a corresponding small binding energy of superconducting electrons, and in turn a lower transition temperature. By this point of view, it seems that the maximum value of $T_c$ in $La_{2-x}M_xCuO_4$ compound should also occur at the doping $x = 0.25$, since the height of potential wells in this case should reach the lowest value allowed by the minimum coherence length. However, the really maximum value of $T_c$ is observed at the doping $x = 0.15$. The possible reason for having this discrepancy may lie in the fact that the minimum coherence length leads the material to have the highest density of superconducting electrons. In this case, when the material is in its superconducting state, the high density of the bound superconducting electrons will make the lattice too hard to compress. That is, the binding energy of superconducting electrons in potential wells cannot monotonously increase with increasing doping, but turns to decrease as the doping x is over a certain value. However, it is obvious that when the doping concentration lies between $x = 0.125$ and $x = 0.25$, there must simultaneously exist two kinds of superconducting chains in $La_{2-x}M_xCuO_4$ compound, in one of them, the superconducting electrons have a coherence length 4a, while, 2a in the other one. We propose that the maximum value of $T_c$ may occur at the superconducting chain with $\xi = 2a$ in $La_{2-x}M_xCuO_4$ compound with the doping $x = 0.15$. In other words, in a normal - doping cuprate compound, the superconducting transition temperature may change from chain to chain, even direction to direction. Since the binding energy of superconducting electrons in these two kinds of superconducting chains is quite different each other, so the superconducting electron can only move in its own chain, cannot transfer between the chains with different coherence length.

However, if the nearest distance between two doped unit cells in cuprates is equal to an odd number times the lattice constant, for instance, $\xi = 3a$, which corresponds to $x = 0.11$, that is, there is one divalent atom in every nine unit cells. In this case, when $t = T/2$, a half of the period of the oscillating supercurrent, the superconducting electrons will arrive at the boundaries between two $CuO_6$ potential wells, where they cannot form a dynamic bound state, so the supercurrent cannot persist, although the superconducting state can be formed. This analysis implies that there must exist a dead point in the superconducting region given in the phase diagrams of cuprate compounds. This phenomenon has been experimentally observed by the scientists at NSLS.[34]

Another important issue which needs to be addressed is that in a superconducting process, the superconducting electrons must coherently move with a lattice deformation wave as shown in the superconducting chains in Fig.6. If an applied electric (or magnetic) field makes the velocity of superconducting electrons much higher than that of the lattice deformation wave, then the phase coherent movement between superconducting electrons and the distortion wave of chain lattice is interrupted, and so the superconducting process is ceased. As is well known, when an applied magnetic field exceeds a certain critical value, the superconductivity in both type I and type II superconductors is destroyed.

On the basis of the mechanism of superconductivity above, it is obvious that if a strain for either compressive or tensile type is introduced into a given superconductor, the superconducting process in the material must be destroyed, since the introduced strain could seriously interrupt the coherent moving of superconducting electrons and lattice deformation waves. In addition, the introduced strain could change the symmetry and volume of the potential wells in the strain area. It follows that the binding energy of superconducting electrons in the strain area must be changed from place to place, so the unique superconducting state cannot be formed in the strain areas. Laan and Ekin have found that just a little strain can cause a large drop in supercurrent of high-$T_c$ superconductors, and suggested that strain effect is intrinsic to the fundamental mechanism of superconductivity.[35]

It is interesting to note that the lattice distortion wave induced by superconducting electrons, unlike the lattice harmonic vibrations, can neither dissipate nor transport energy by themselves, since superconducting electrons must keep periodically exchanging their excitation energy with chain lattice. The time variation of the energy stored in superconducting electrons and chain lattice can be expressed as

$$E_e(t) = E_m \sin^2(\omega t) \quad \text{for superconducting electrons and}$$

$$E_l(t) = E_m \cos^2(\omega t) \quad \text{for a chain lattice} \tag{8}$$

Where $\omega = 2\pi v_s/\lambda$, the angular frequency, $v_s$ is drift velocity of superconducting electrons, $\lambda$ is the wave length of oscillating supercurrent, or the wave length of lattice distortion wave, which is also equal to the coherence length in the supercurrent direction, $E_m = \Delta V \times e$, the energy gained by a superconducting electron from applied dc voltage source, or the excitation energy of a superconducting electron. If the applied field is in [100] and [110] directions of La$_{1.875}$M$_{0.125}$CuO$_4$ compound, then the wave length of oscillating supercurrent corresponds to $\lambda = $ 1.6 nm and 1.1 nm, respectively as shown in Fig. 5. Apparently, a superconductor itself is a microwave generator with a wide tunable frequency range by adjusting the applied voltage. However, we can see from the discussion above that a superconductor by itself cannot make any electromagnetic emission, although there does exist oscillating supercurrent in it. Following the discussion above, it is easy to see that the tunable frequency for La$_{2-x}$M$_x$CuO$_4$ should fall in range of

$$(1 - a/\xi) < f T_d < (1 + a/\xi) \qquad (9)$$

Where $T_d$ is period of a lattice distortion wave, a and f represent respectively the width of potential wells and the frequency of oscillating supercurrent in a given material. As we have noted that the coherence length in cuprate superconductors must satisfy $\xi = 2na$. Suppose that $\xi = 2a$, the minimum coherence length, then the tunable frequency range will fall in $0.5 < f T_d < 1.5$. However, if $\xi = 4a$, like the case in the underdoped cuprates, then tunable frequency will fall in the range $0.83 < fT_d < 1.16$. Thereby, the tunable frequency range in a given superconductor varies with coherence length. Thus we come to a conclusion that for cuprate superconductors, the smaller the coherence length, the larger the tunable frequency range.

In the BCS model, a superconductor at below $T_c$ has an energy gap which is equal to the energy needed to break apart one of the Cooper pairs. When the temperature is raised above $T_c$, the energy gap completely disappears. However, the energy gap in cuprates varies with temperature in rather different way. At below $T_c$, the superconducting gap changes with direction in the momentum space, while at above $T_c$, the energy gap does not go away. These two phenomena are usually referred to as pseudogap.[36] Obviously, this so - called pseudogap behavior observed in cuprate compounds raises an even bigger challenge to the old theory, but it is quite reasonable from the point of view of this new model proposed in this paper. As has been pointed out, the height of the CuO$_6$ and CuO$_5$ potential wells in cuprates changes with crystal direction in the real space, so it must follow that the binding energy of the superconducting electrons in potential wells also changes with direction in the momentum space. In addition, it is clear that no matter how many bound superconducting electrons are contained in a given superconducting chain, as long as one superconducting electron in the chain is excited from its bound state into the free electronic states by thermal energy, then the superconducting process in the chain is totally suppressed. So it is not surprising that when the cuprates are heated up above $T_c$, there may still have a portion of superconducting electrons staying in their dynamic bound state, perhaps some of them may stay in their bound state even at room temperature. For this reason, it is quite difficult to find a unified formula that can uniquely predict the transition temperature for all superconductors.

Both x - ray diffraction data and neutron scattering data on $La_{2-x}M_xCuO_4$ compound have identified that when the compound makes a transition from normal metallic state to superconducting state, the lattice structure undergoes an order - disorder transition due to the incommensurate displacements of atoms from their ideal crystallographic sites.[37,38] This behavior has been observed in all high- $T_c$ superconductors. In macroscopic theory of superconductivity, it seems reasonable from point of view of thermodynamic theory that the transition from normal metallic state to superconducting state was called as superconducting phase transition. But according the model we give above, the order - disorder transition accompanied by the superconducting state, in fact, is resulted from the lattice distortion induced by dynamic bound state of superconducting electrons in the same crystalline phase, as shown in Fig. 5. Actually, the neutron scattering data on $La_{2-x}M_xCuO_4$ shows clear tetragonal pair peaks, but which were broadened by random displacements of atoms from their ideal sites.[39] Therefore, we can draw the conclusion that the onset of superconductivity must be accompanied by lattice distortion driven by the dynamic bound state of the superconducting electrons. Both the normal state and superconducting state for a given superconductor must exist in the same crystalline phase, no matter how big the dynamic distortion could be.

We have mentioned that in the normal state of $La_{2-x}M_xCuO_4$ compound, only the La (5d) electrons in doped unit cells have a chance to become free electrons at the energy levels above $CuO_6$ potential wells and conduct current through $CuO_2$ planes. In this case, the high-energy La (5d) electrons release their energies into entire volume of the cuprate compound, which will make the entire lattice relatively soft compared to that of the pure La214. At x = 0.125, the density of high- energy electrons in $La_{2-x}M_xCuO_4$ compound is about $1.6 \times 10^{20}/cm^3$ which by equation (6) will lead to a decrease of about 5 GPa in lattice modulus. Since the velocity of sound in a medium is proportional to the square root of the bulk modulus of the material, thus, the sound velocity in the metallic phase of $La_{2-x}M_xCuO_4$ compound should show a decrease compared to that in the pure La214. However, when the compound transits into superconducting state at below $T_c$, the superconducting electrons become dynamic bound electrons in doped cells, which, in turn, will give rise to an increase of about 25GPa in the lattice modulus compared to that in the pure La214. This result means that the lattice in cuprates will become more hardening in its superconducting state compared to that in its normal state.[40] Now we can see why the sound velocity measured by the vibrating reed method in superconducting state of cuprates is greatly increased compared with that measured in its normal state. Based on the anomalous increase of the sound velocity in cuprates at temperature below $T_c$, Bishop's group has proposed that the superconducting process in cuprates must result from the electrons at the energy state far from Fermi surface.[41,42]

However, the situation is quite different for conventional superconductors. The potential wells for trapping superconducting electrons in most of conventional superconductors come from crystalline grains. They usually have a volume as large as five to ten orders of magnitude larger than that in cuprates, and so the negative pressure induced by a superconducting electron in a given crystalline grain cannot exert a strong compression on its neighbors. Thus, the negative pressure in the crystalline grains occupied by superconducting electrons should make the lattice of conventional superconductors softening in their superconducting state compared to that in their normal state. Here we need to point out that even in some high $-T_c$ superconductors, as long as the connections between the potential wells for trapping superconducting electrons have a weak interaction, as in the case of $A_3C_{60}$ fullerides, the lattice in the area occupied by the superconducting electrons should become softening in superconducting state.

Another fundamental character that needs to be discussed is the effect of replacement of Cu by Ni and Zn atoms on the properties of $La_{2-x}M_xCuO_4$ superconductors. From the superconducting model illustrated above, it is obvious that the binding energy and symmetry of the dynamic bound state of superconducting electrons in a given superconducting chain must remain exactly the same in every potential well, otherwise the scattering centers for destroying supercurrent will be introduced. If a copper atom is replaced by Ni or Zn atom in a given octahedron, then the highest unoccupied states in Ni $(Zn)O_6$ potential well should be Ni (Zn) 4s - O2p$\sigma$ antibonding states. Thus the binding energy and symmetry of the superconducting electron in the Ni $(Zn)O_6$ octahedron must be quite different from that in other $CuO_6$ octahedrons. Obviously, if the doped Ni $(Zn)O_6$ unit cell appears in a superconducting chain, then this doped cell will become strong scattering centers for superconducting electrons, and so superconducting process cannot persist. So we can say that any isovalence element, especially those which have strong atomic magnetic moment, replace Cu atom in high – $T_c$ cuprates, the superconductivity in the doped compound must be seriously suppressed with increasing doping.[43] For the same reason, it is not surprising that the fully metallic cuprate compounds $La_4Ba\,Cu_5O_{13}$ and $La_5SrCu_6O_{15}$ do not show any bulk superconductivity. Since the lattice structures for both compounds are constructed by alternating arrangement of copper octahedrons and copper pyramids, thus, the unique superconducting state cannot be formed in this sort of compounds.[44]

Here we need to point out that the BaK( Bi,Pb )$O_3$ compound, unlike $BaKBiO_3$, does not belong in the group of high- $T_c$ superconductors. In this compound, the potential wells for trapping superconducting electrons consist of both $BiO_6$ and $PbO_6$ octahedrons. The binding energy of superconducting electrons in these two kinds of octahedral potential wells is undoubtedly different. So the superconducting process cannot be started, even there do exist the dynamic bound states in both kinds of potential wells.

In addition, $Nd_{2-x}Ce_xCuO_4$ is a typical electron - doping cuprate compound, which has a lattice structure similar to that of La214 compound. The difference between the two lies in the position of the $O_z$ atoms. In the lattice structure of $Nd_{2-x}Ce_xCuO_4$, the two $O_z$ atoms are displaced from their apex positions to the sites on the faces of the tetragonal cell. So in $Nd_{2-x}Ce_xCuO_4$ compounds, the Cu ions are in square planar coordinated with oxygen atoms, there is no three-dimensional potential well surrounding Cu ions, like $CuO_5$ and $CuO_6$ in the hole - doping cuprates. So both the electron - doping cuprates and BaK( Bi Pb)$O_3$ belong in the regime of conventional superconductors. That is, the potential wells for trapping superconducting electrons in both BaK( Bi,Pb )$O_3$ and electron - doping compounds are not resulting from lattice structure itself, but from the microstructures, like superlattice or crystal grains. The large and uncertain volumes of the potential wells make both compounds have a relatively low and sample - dependent transition temperature.[45]

Now we are at the position to discuss the effect of replacement of La or Y ions by the trivalent rare - earth elements on the superconducting properties of the corresponding compounds. It is well known that both La and Y in their corresponding compound La214 and YBCO are located in charge reservoir area, which is isolated from superconducting chains in both compounds. So the replacement of La and Y by the trivalent rare - earth elements, the superconducting properties in both compounds should not be affected by the magnetic moments arising from the incomplete 4f shell of these ions. As we have emphasized above, the primary function of La and Y ions in their corresponding compound is to provide superconducting electrons by their d shell electrons. According to CIDSE model, the binding energy of superconducting electrons is directly related to the energy level of La 5d electron and Y 4d electron in their corresponding compound. One criterion for superconductivity is that every superconducting electron in a superconducting chain must have the same binding energy. Otherwise, the scattering centers for superconducting electrons will be unavoidably introduced. For this reason, we can expect that by using any trivalent rare - earth element to partially replace La or Y atoms in their corresponding compounds, the superconductivity in the ordinary compound must be suppressed, unless the binding energy induced by the 4f electron of the doping element has exactly the same as that of the original one. It has been found that even one percent of Y ions in YBa$_2$Cu$_3$O$_7$ are replaced by Gd, the superconductivity is completely destroyed in the doped compound. However, if the Y atoms are completely replaced by trivalent rare - earth elements, then it can be expected that compounds REBa$_2$CuO$_7$ (RE = Nd, Sm, Gd, and Eu) should exhibit stable bulk superconductivity, in spite of the big magnetic moment on the rare - earth sites. The transition temperature in REBa$_2$CuO$_7$ should depend upon the energy level of 4f electrons in the corresponding compound. Since the ionization energy of the 4f electron in rare - earth elements does not show remarkable changes (the ionization energy for element Nd, Sm, Eu, and Gd is 5.51, 5.6, 5.67 and 6.16 eV, respectively), so the transition temperature observed experimentally in the REM$_2$CuO$_7$ compounds varies slightly from 85 K to 100K (from Ref. 46, 47). However, it is important to note that in the entire lanthanide series (RE)$_2$CuO$_4$, only with RE = La is Cu weakly octahedrally coordinated with oxygen ions, which make La$_2$CuO$_4$ have a chance to become a high - T$_c$ superconductors. For all the other (RE)$_2$CuO$_4$ compounds, Cu is in square- planar coordinated with oxygen ions.[16] None of them can be doped to be high–T$_c$ superconductors. Perhaps the typical example in (RE)$_2$CuO$_4$ compounds is to find the electron - doping superconductor Nd$_{2-x}$Ce$_x$CuO$_4$. This fact further proves that the three - dimensional potential wells formed by lattice structure itself is a key criterion for achieving a high–T$_c$ superconductor.

Perhaps the most important function of oxygen in high-$T_c$ cuprates is to provide the three-dimensional potential wells surrounding Cu ions (octahedral or pyramidal coordinated), into which the superconducting electrons trap themselves to become a dynamic bound state. Thus, the oxygen content must play a dramatic effect on the superconducting properties of the cuprate compounds. In fact, metallic conductivity and superconductivity have been achieved by increasing the concentration of oxygen over the stoichiometric value in $La_2CuO_{4+y}$ compound.

It has been observed that the extra oxygen atoms occupy interstitial positions in layers of La-O (Ref. 23). In this case, the function of extra oxygen, similar to that of divalent element doping, is to absorb the extra 5d electrons of La atoms in the cells near to the doping oxygen. Since in stoichiometric $La_2CuO_4$ compound there are two La 5d electrons in each unit cell, it is impossible for two high-energy electrons to simultaneously form a dynamic bound state in the same $CuO_6$ potential well. So that no matter what method that might be used to synthesize La214 to be a superconductor, as long as a 5d electron of La in certain cells can be removed, and the concentration of the cells with one 5d electron satisfies the requirement of coherence lengths, the superconductivity can be achieved.

In addition, it has been found that by removing some of $O_z$ atoms from $La_{2-x}Ba_xCuO_4$ compound, the transition temperature above 50K was observed in small islands.[48] This is not surprising from the point of view of the CIDSE model, removing a certain amount of $O_z$ atoms from La214 compound should not have important effect on both the energy state of La 5d electron and the height of potential wells in $CuO_2$ plane direction, but can change the type of potential well from octahedron to pyramid. Since the volume of a $CuO_5$ pyramid potential well is roughly equal to one half of that of the $CuO_6$ octahedron, so one can expect that the binding energy of superconducting electrons in a pyramid potential well should roughly be $\sqrt{2}$ times that occurred in $CuO_6$ octahedral potential wells. Suppose that the superconducting transition $T_c$ is proportional to the binding energy of superconducting electrons, then the transition temperature in this case should be close to $\sqrt{2} \times 40\ K = 56\ K$.

Perhaps the most complex and interesting issue about oxygen doping is that the physical properties of $YBa_2Cu_3O_{7-y}$ based compounds are dramatically affected by oxygen content. For instance, at y = 0, the compound $YBa_2Cu_3O_7$ has a metallic phase with a transition $T_c$ = 90 K. However, as oxygen content is decreased to y = 0.6, the compound transforms into the semiconducting phase with a tetragonal symmetry. The band structure of $YBa_2Cu_3O_m$, m = 6, 7, and 8 have been calculated with a variety of methods. The band structures resulting from different methods all consistently demonstrate that $YBa_2Cu_3O_m$ compounds with either tetragonal or orthorhombic symmetry all have metallic character.[2,49] The discrepancy between experimental results and theoretical predictions on these compounds is essentially similar to that happened in $La_{2-x}M_xCuO_4$ compound. The primary reason to lead this discrepancy for all cuprates lies in the fact that the Cu-O pdσ antibonding state (primarily consisting of Cu $d_{x^2-y^2}$ orbitals), which is the only one to conduct ballistic metallic current, is confined by $CuO_6$ or $CuO_5$ potential wells in the corresponding compounds. In $YBa_2Cu_3O_7$ compound, the superconducting electrons are provided by Y 4d electrons, which are lying at about 5 eV in energy above Fermi level calculated from the data of inverse photoemission spectrum.[50] On the other hand, the band structure of $YBa_2Cu_3O_7$ shows that the maximum of the Cu-O pdσ antibonding band is about 2 eV above Fermi level.[51] While an insulator sample with y = 0.8 shows an optical gap about 1.75 eV measured by optical conductivity.[23,52] So it is reasonable to assume that the height of $CuO_5$ potential well in $YBa_2Cu_3O_7$ should be lower in energy than 1.75 eV. Thus, the normal state of $YBa_2Cu_3O_7$ compound has metallic character. It has been found that by heating the YBCO above 500° C, the oxygen in Cu-O chain structure will first leave the sample since the Cu-O bond in the chain structure is relatively weaker compared with that in $CuO_2$ planes. While with removing oxygen atoms out of the compound from the chain structure, the electric charge is automatically transferred from chain to plane,[23,53] which in turn increases the height of $CuO_5$ potential well along the plane direction. It is apparent that the energy state of Y 4d electron cannot be affected by the change of oxygen content in chain structure. Following such a upward shift of potential height, the transition energy from Y 4d level to the level $E_b$ of potential wells continuously decrease with removing oxygen out of the compound, which in turn leads to a decrease in the binding energy of superconducting electrons. Now we can understand the well known puzzling phenomenon why the superconducting transition temperature in $YBa_2Cu_3O_{7-y}$ smoothly decreases with removing oxygen atoms out of the compound. When the oxygen content in $YBa_2Cu_3O_{7-y}$ compound is reduced to y = 1, the electric charges transferred from chain to plane are enough to raise the height of $CuO_5$ potential well over the maximum of Cu - O pdσ antibonding band. Then superconductivity disappears and the compound becomes an insulator without an energy gap.

It is important to realize that the band structures of $YBa_2Cu_3O_7$ calculated by several groups all consistently show that both chain and layer exhibit metallic character. However, a number of experiments, such as oxygen content, substituting Cu by Fe or Zn, all demonstrate that the main role in the appearance of superconductivity in $YBa_2Cu_3O_7$ compound is played by the $CuO_2$ planes.[54] The most remarkable difference between chain and plane lattice structures is that the Cu ions in planes have three - dimensional coordinated $CuO_5$ pyramids, while the Cu ions in chains are in square planar coordinated. This fact further demonstrates that the stable dynamic bound state for superconducting electrons can only be formed in three- dimensional potential wells. Compared with $La_{2-x}M_xCuO_4$ compound, the volume of $CuO_5$ potential well in $YBa_2Cu_3O_7$ roughly is a half of that of the former. While the energy using to create superconducting bound state in $YBa_2Cu_3O_7$ is about 2.5 times that occurred in $La_{2-x}M_xCuO_4$. Thus the superconducting transition $T_c$ in $YBa_2Cu_3O_7$ should be close to $(\sqrt{2} \times \sqrt{2.5} \times 40\,K) = 89\,K$.

A correct model of superconductivity should consistently explain the complex phenomena observed in all cuprates. Another well known puzzling phenomenon is that the transition temperature increases as we move from single plane cuprates to those containing two and three $CuO_2$ planes in their unit cells. The Bi-based and Tl - based compounds have the general formula $A_2Ca_{n-1}B_2Cu_nO_{2n+4}$, where A = Bi (Tl) and B = Sr (Ba). There exist three sorts of copper oxygen layers in their structures, which with increasing structure index n are the single octahedral layer for n = 1, and double $CuO_5$ pyramidal layers separated by $Ca^{2+}$ ions for n = 2, as well as two $CuO_5$ pyramidal layers and one square planar layer all separated each other by $Ca^{2+}$ ions for n = 3 (Ref. 55). As mentioned before, the superconducting dynamic bound state cannot be formed in the two - dimensional $CuO_2$ square plane, so the $CuO_2$ planes in Bi- based compounds with n > 3 have no direct contribution to superconductivity. The superconducting layers, that is, the octahedral and pyramidal layers in Bi and Tl - based compounds are basically similar to that of LBCO and YBCO compounds, except that the Cu-O distance along the c axis is 2.6 Å (2.7 Å) for Bi (Tl) compounds, respectively. Since the electronic structure and superconducting properties in both compounds are considerable resemblance, in the following, we will only take the Bi- based compound as a typical example to further test the general superconducting mechanism we proposed above.

As in the LMCO and YBCO systems, the superconducting properties in Bi (Tl) -based compounds are primarily dominated by the pdσ antibonding band derived from superconducting layers ($CuO_6$ and $CuO_5$ layers). The band structure of $Bi_2Sr_2CaCu_2O_8$ calculated by Krakauer and Pickett shows that the pdσ antibonding band from Cu- O1 layers crosses $E_F$ and reaches a maximum of about 2.4 eV above $E_F$, estimated from the same band structure above. It is reasonable to assume that the height of $CuO_5$ potential well in Bi -based compounds should be essentially similar to that in YBCO compound. So the normal states of Bi - based compounds show a metallic character in the superconducting planes. In addition, two Bi - O layers also provide two widely dispersive bands derived from the Bi 6s - O2p antibonding bands, which crosses $E_F$ and reaches a maximum of about 2.2 eV above $E_F$ estimated from the band structure in Ref. 51. This character of Bi - O band indicates that the Bi - O layers lying between $CuO_2$ layers are also in metallic state.[56] However, the formal valence arguments have assigned a 3+ state to the Bi ions in the Bi - based compounds. Thus the oxygen ions surrounding Bi ions should be close to 2 - state, so that, the height of $BiO_6$ potential wells should be much higher in energy than that of the maximum of Bi 6s – O2p antibonding band. For this reason, the Bi- O layers in Bi - based compounds manifest themselves like an insulator. Even so, the Bi- O layers still play an important role in the onset of superconductivity in these Bi - based compounds. The $BiO_6$ octahedral potential wells provide an electron reservoir, which can automatically regulate the density of high - energy electrons to satisfy the requirement for forming a superconducting state.[57] That is, the function of the Bi-O layers in Bi - based compound is similar to the divalent atoms doping in the $La_{2-x}M_xCuO_4$ system.

It has been found experimentally that the transition temperature $T_c$ in Bi - based compound depends on the number of $CuO_2$ planes in a unit cell and shows the values 10, 85 and 110 K for index n = 1, 2, 3, .., respectively. Following this new model proposed above, the tendency of superconducting transition $T_c$ in Bi - based compounds is not surprised. For n = 1, the $Bi_2Sr_2CuO_6$ (B2201) compound, there only exists a single $CuO_2$ layer per unit cell, in which Cu ion is octahedrally coordinated with oxygen, similar to the case in La214 compound. In Bi-based compound, the superconducting electron is provided by Sr 5s electrons. Since the ionization energy of Sr 5s is lower than that of La 5d electron, it has been found that when the divalent atom Sr in B2201 compound is replaced by La atoms with x = 0.35, the transition temperature in $Bi_2Sr_{1.65}La_{0.35}CuO_6$ can be reached to 29K close to that obtained in $La_{2-x}M_xCuO_4$ (Ref. 58).

For n = 2 and n = 3 compounds Bi2212 and Bi2223, they have 2 and 3 CuO layers all separated by $Ca^{2+}$ ions in their unit cell, respectively. It has been suggested that $Ca^{2+}$ ions provide one electron to the CuSrBiO slab on either side.[2] In both Bi2212 and Bi 2223 compounds, the potential wells for trapping superconducting electrons are provided by $CuO_2$ pyramids, whose volume is just a half of that of the octahedron in Bi2201 compound. This means that the $T_c$ in both Bi2212 and Bi2223 compounds should roughly be $\sqrt{2}$ times that occurred in Bi 2201 compound. Moreover, perhaps the more important factor that makes $T_c$ increase in Bi - compounds, lies in the increase in energy of Sr 5s electrons with increasing the index number n due to the internal stress caused by getting close together of $CuO_2$ planes. According to the combined direct and inverse photoemission data for $Bi_2Sr_2CaCu_2O_8$, the Sr 4d states show a broad spectral peak about10 eV width above $E_F$ level.[29] If we presume that the state of Sr 5s has the same energy region as that of Sr 4d state and take the energy level of Sr 5s state to be 8 eV above $E_F$, then the $T_c$ in $Bi_2Sr_2CaCu_2O_8$ compound is about 100 K estimated by equation (6).

Another remarkable feature of Bi-based compounds is that the lattice distortion not only presents in the $CuO_2$ planes like observed in other cuprate compounds, but also occurs in Bi -O layers, which cannot be simply explained by mismatch of ionic radii. As mentioned above, in all octahedrally and pyramidally coordinated cuprate compounds, the doping hole concentration required by coherence lengths should be around at 0.0625, 0.125 and 0.25 per unit cell. When the doping hole concentration has a value of 0.111 per unit cell, the superconducting process is totally suppressed. While in Bi - based compounds, the more than 50% high - energy electrons must be accepted by $BiO_6$ structure. So it is not surprising to observe the large distortions in Bi-O and Sr - O planes.

Finally, we need to point out that the superconducting transition temperature measured for Tl - based compounds has the value 85, 105, and 125 K for n = 1, 2, 3, respectively. Based on the CIDSE model, we can assert that the superconducting process for Tl - compounds with n = 1, like in the case for n = 2 and 3, should also occurs in $CuO_5$ pyramidal layers. This situation perhaps is caused by the strong distortion of oxygen atoms along the c axis near the transition temperature.[59]

## IV. ON JOSEPHSON EFFECT AND THE UNIT OF MAGNETIC FLUX QUANTIZATION

In 1962, Brain Josephson predicted that when two pieces of superconductor are separated by a thin oxide layer, the Cooper pairs can tunnel through such junctions without any resistance, and if a DC voltage V is applied across the Josephson junction, an alternate current is produced with a frequency f = 2eV/h. This proposal is very puzzling in physics, but the Josephson equation is so accurate that physicists can measure a fundamental constant e /h to unprecedented precision. This undoubted fact makes physicists absolutely accept that Cooper pair is the only mechanism responsible for superconductivity. However, the physics behind Josephson tunneling has never been consistently understood in the BCS theory. For instance, in SIN (superconductor - insulator - normal metal) tunneling, the threshold voltage for tunneling is $\Delta/e$, which in BCS theory is referred to as single particle tunneling, where $\Delta$ represents one half of the energy gap of the superconductor. While in SIS tunneling system, there is the same superconductor on both sides of the barrier, the threshold voltage in this case is $2\Delta/e$, which is identified as Cooper pairs tunneling through insulating barriers. However, it has been found experimentally that if n superconducting junctions are coupled together in series, then the threshold voltage is $n\Delta/e$. Based on the experimental results from a single junction to multijunctions, the only conclusion which we can draw is that the single particle tunnels through n junctions, and the voltage drop across each piece of superconductor is $\Delta/e$. Now the question is raised, does the factor of 2 in Josephson tunneling equation derive from SIS system represent 2 electrons or the Cooper pair? If the answer is yes, then the conclusion is in contradiction with the results obtained from multijunction systems. However, if the factor of 2 in Josephson equation is arising from the voltage drops across two pieces of superconductor, then the results obtained from the entire tunneling systems lie in the consistency. In addition, if a SIS junction is consisted of two different superconductors, which have a half of the energy gap $\Delta_1$ and $\Delta_2$ respectively, then the threshold for the tunneling voltage is identified experimentally as $(\Delta_1 + \Delta_2)/e$, which also demonstrate that the voltage drop across two the pieces of superconductors on both sides of a junction is $\Delta_1 / e$ and $\Delta_2 / e$, respectively.

Undoubtedly, the $\Delta/e$ voltage drop across each piece of superconductor definitely cannot be explained in the BCS theory. But it is a natural result of the superconductivity mechanism proposed here. In this new model, superconducting state consists of the dynamic bound state of superconducting electrons in three - dimensional potential wells lying at above the Fermi level. The thermal excitation energy of a superconducting electron, $\Delta$, which is defined as one - half of energy gap in BCS model, is exactly equal to the binding energy of superconducting electrons in their dynamic bound state.

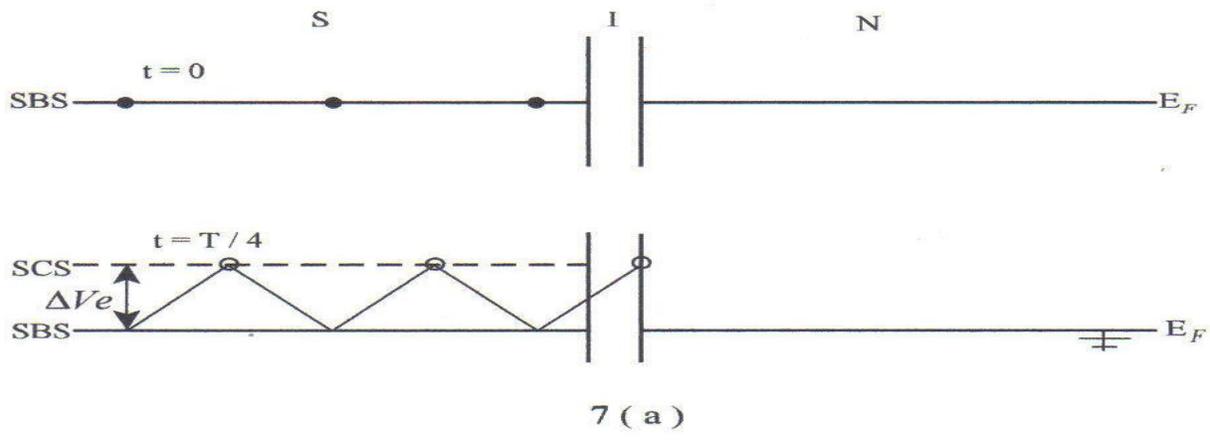

7 ( a )

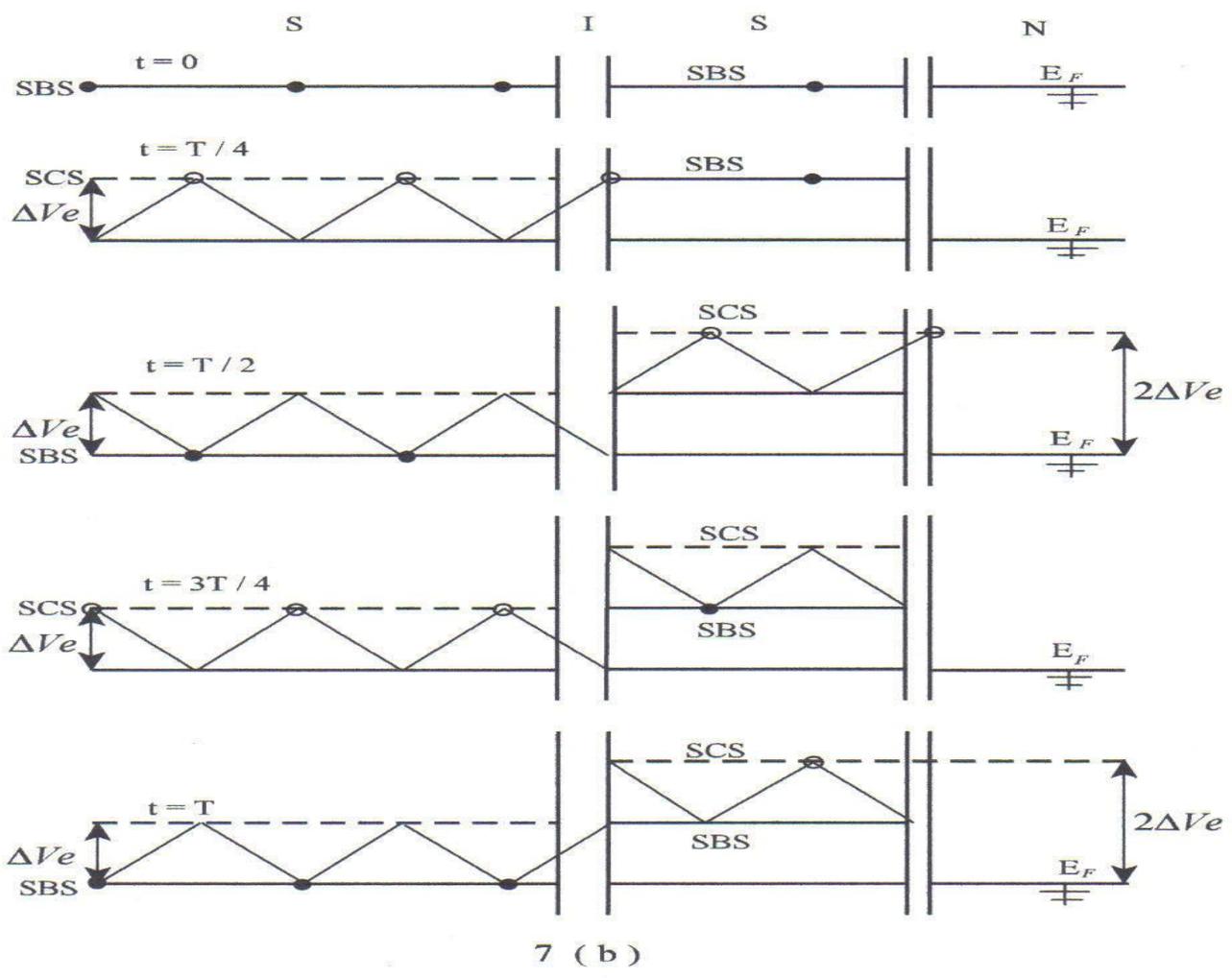

7 ( b )

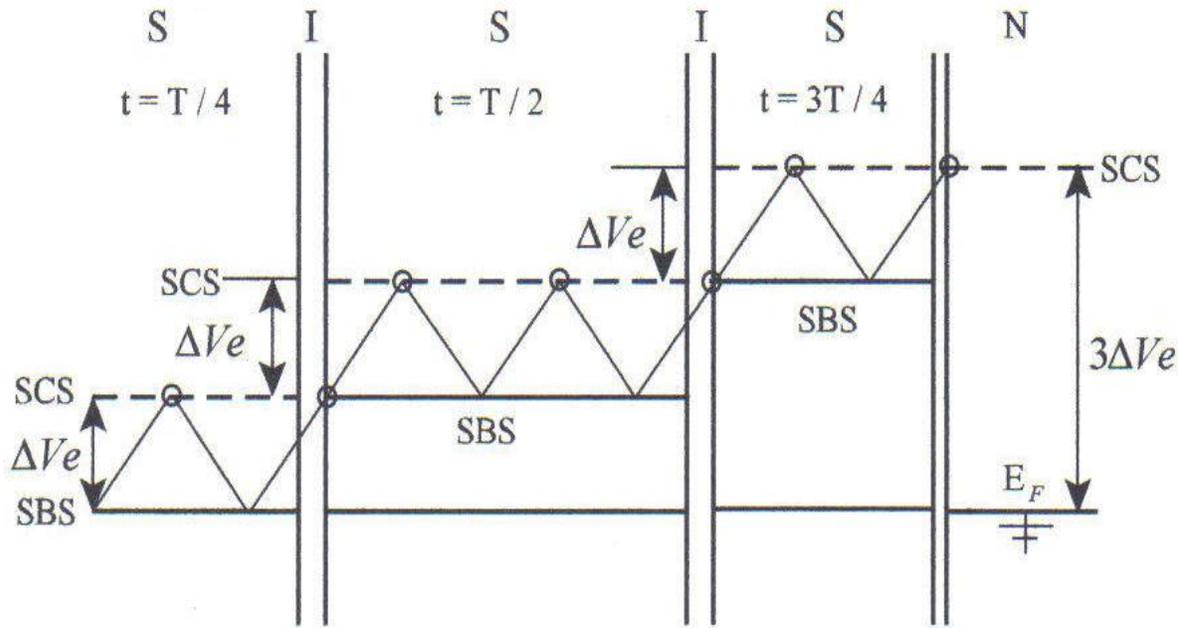

FIG. 7. The Josephson tunneling processes. The diagram (a), (b) and (c) show the superconducting elecreon tunneling processes in superconductor - insulator - normal metal (SIN), sperconductor - insulator - superconductor (SIS), and SISIS systems. The left - hand side of barrier for the three systems is connected to the negative terminal of a DC voltage source, while the right - hand side for the three systems is all connected to a grounded metal electrode. The solid and open circles denote the superconducting electron in its dynamic bound state (SBS) and conducting state (SCS), respectively. The tooth - wave curves represent symbolically the variations of the energy of superconducting electrons with lattice distortion waves

In Fig. 7, we schematically show the Josephson tunneling processes in SIN, SIS and SISIS systems, respectively. The solid circles denote the superconducting electrons at their dynamic bound state, and the open circles represent superconducting electrons in conducting state. The distance between the nearest two solid circles (or open circles) is equal to the coherence length or wave length of superconducting electrons, which is also equal to the wave length of the lattice - distortion wave. The tooth-wave curves represent the variations for both the energy of superconducting electrons and lattice distortion waves with time and position. We assume that the left- hand side of the barrier for the three systems is connected to the negative terminal of a dc voltage source, and the right - hand side of barrier for three systems is all grounded. In this superconductivity model, at below $T_c$ the highest occupied state in a superconductor is the dynamic bound state of superconducting electrons in potential wells lying at above the Fermi level of the material. Physically, it is reasonable to take the energy level of dynamic bound state of superconducting electrons as the quasi - Fermi level of superconductors.

In the SIN tunneling system, at zero voltage and below $T_c$ the Fermi levels on both sides of the junction stay in equal as shown in Fig. 7a. When the negative voltage drop across the superconductor is lower than the threshold value $\Delta/e$, there is no current flowing through the junction. As the applied negative voltage - $\Delta V$ exceeds the threshold value $\Delta/e$, that is, $(-V) \times (-e) > \Delta$, the bound superconducting electrons become free electrons and begin to drift under the applied field. The superconducting electron nearest to the junction will first tunnel through the insulator barrier into metal section, and then the tunneling current sharply occurs between the superconductor and normal metal. In this case, the tunneling process is achieved by superconducting electrons tunneling through the junction, and the junction does not show any resistance. In contrast, suppose that the metal section on the right - hand side of the barrier is connected to the negative terminal of the dc voltage source, while the superconductor on the left - hand side of the barrier is grounded. In this case, the negative voltage drop on metal section will raise the energy of free electrons from $E_F$ with a scale as $(-\Delta V) \times (-e)$. When the energy of free electrons in metal section is lower than $\Delta$, the free electrons cannot tunnel through the barrier into the superconductor, because there are no free - electron states in the superconductor. As the energy of free electrons in metal is over $\Delta$, the normal electrons begin to tunnel from metal into the conducting states of the superconductor, and an ohmic tunneling current will be measured in the external circuit. We predict that the behavior of the tunneling current in this case should be quite different from that occurred in the former, and to further study this different behavior may favorably reveal a real mechanism hidden behind the superconducting tunneling phenomenon.

In order to illustrate the tunneling processes in SIS system more clearly, in Fig. 7(b), the five panels from $t = 0$ to $t = T$ are used to express the time variations of the energy and positions of superconducting electrons. At $t = 0$, in thermodynamic equilibrium and below $T_c$, the quasi - Fermi level in the two superconductors on both sides of junction stays in equal. When the voltage drop across the left - hand side superconductor satisfies $(-\Delta V) \times (-e) > \Delta$, all superconducting electrons in this piece of superconductor become free electrons and begin to move toward to the junction. After $t > T/4$, the superconducting electron nearest to the junction begin to tunnel through the junction, and at the same time, the superconductor at the right - hand side of the junction feels a voltage - $\Delta V$, which makes the entire energy structure of the right - hand side superconductor shift up by $(-\Delta V) \times (-e)$ as shown in Fig 7(b). Then the external dc voltage source starts making the voltage drop across the right - side superconductor. When the condition $(-\Delta V) \times (-e) > \Delta$ is reached on the right - hand side superconductor, the entire SIS system becomes superconducting without any resistance. It can be easily seen from the tunneling panels that the phase difference between the two superconductors is $\pi/2$, and in each oscillation period of superconducting electrons (or distortion waves of chain lattice), there is one superconducting electron tunneling through the junction in each superconducting chain. As we have noted above, the oscillating supercurrent flowing within a superconductor cannot emit electromagnetic waves, since superconducting electrons must keep periodically exchanging their excitation energy with a chain lattice to maintain the supercurrent in constant. Only the superconducting electron that moves out of a superconductor has a chance to transfer its energy gained from an external voltage source into the electromagnetic radiation waves. For this reason, the superconductor on the right - hand side of the barrier needs to be connected to a grounded metal electrode as shown in Fig. 7(b). Since the highest occupied state in a superconductor at below $T_c$ is the dynamic bound state of superconducting electrons, there are no free electrons available for achieving an ohmic contact, so no matter what material is placed to contact with superconductors, there always exhibits a Josephson junction. When a superconducting electron tunneling into the metal electrode on the right - hand side of junction falls back to the ground, the electron has a chance to emit an electromagnetic wave with a frequency $f = 2e\Delta V/h$, which in form is quite similar to the well-known Josephson equation derived on the basis of Cooper pair. The difference is that the $\Delta V$ in the equation above represents the voltage drop on each piece of superconductors in SIS system, while the voltage V in Josephson equation was defined as the voltage drop across the junction. Following the tunneling processes in SIS system, it is straightforward to find that in the SISIS system, the radiation frequency emitted by a superconducting electron tunneling into the metal electrode should be $3e\Delta V/h$ as shown in Fig. 7(c). For the same reason, if there are n junctions coupled together in series, the emission frequency from the last junction will be $(n + 1)e\Delta V/h$, assume that the system is fabricated from the same superconductor. Obviously, if a multi - junction system is fabricated from different superconductors, then the final emitting frequency generated by the last junction should be equal to $\sum_i e \Delta V_i / h$, summing the voltage drops over all superconductor sections. We therefore draw a conclusion that in a Josephson tunneling system, the frequency of oscillating current in each superconductor section has a value $\Delta/h$, where $\Delta$ is the excitation energy of superconducting electrons in the corresponding superconductor. The junction itself is not an energy source for generating radiation emission. The function of the insulating barriers in multijunction systems is to coherently add the excitation energy of superconducting electrons in each superconductor section together.

It is important to note that the tunneling equations given above cannot hold, until the coherent tunneling condition is achieved. The optimal tunneling configuration for the Josephson tunneling system should satisfy the following conditions. In the superconductor on the left - hand side of the junction [see Fig. 7], the distance from junction to the nearest bound superconducting electron should be smaller than one half of coherence length, and on the right - hand side of the junction, the insulator layer should locate at the position of the first bound electron site, so that this bound state is empty, just like the configurations shown in tunneling panels in Fig. 7. As the junction shifts from its optimal position the tunneling current will sharply decrease. Obviously, the optimal tunneling configuration given above is only for the case that the superconducting electrons tunnel through a junction from the left - hand side of the junction. On the contrary, if the superconducting electrons tunnel through the barrier from right - hand side of the junction, this configuration will totally destroy the tunneling process. Thus we can see that the position of junction plays a crucial important role in the superconducting tunneling processes. Here we need to emphasis that even if the superconducting electrons tunnel into the grounded metal electrode at the optimal tunneling condition, it is still impossible to get the expected radiation frequency from a given junction, since the continuous levels of the conduction band in the metal electrode are favorable to transfer the transition energy into thermal energy. For this reason, historically, the Josephson effect cannot be directly proven by measuring the radiation frequency from a tunneling junction biased under a dc voltage source.

In modern physics, it is widely accepted that tunneling is a quantum phenomenon, which is arisen from the fact that the wave - function describing a tunneling electron can partially penetrate a thin insulating barrier. This mechanism perhaps works for explaining the tunneling behavior between normal metals. But it is not enough for explaining the tunneling phenomena occurred between superconductors. As noted above, in a superconducting process, the superconducting electrons move coherently with a lattice distortion wave, just like the case illustrated by both the superconducting chain in Fig. 6 and the tunneling panels in Fig. 7. Therefore, we can imagine that in a superconducting process, the superconducting electron is riding on the lattice distortion wave and moves forwards together with it. If the width of the insulator layer is thin enough compared with the wavelength of the lattice distortion wave or the coherence length, then the thin insulator layer could not seriously disturb the lattice distortion wave, so that the superconducting electron riding on the lattice distortion waves can pass through the insulator layer with the lattice distortion wave together without any scattering. However, when a Josephson junction biased under a dc voltage is subjected to an electromagnetic field, the coherent movement of superconducting electrons and lattice distortion waves must be to some extent destroyed by the applied electromagnetic field. In this case, the sharp onset of tunneling current disappears and the resistivity cannot be extremely small at all temperature below $T_c$.

Thus, we propose that tunneling is a phenomenon caused by the coherent movement of the electrons and lattice distortion waves. As long as the electron has enough energy to excite a lattice distortion wave and moves coherently with it, then the electron moving coherently with a lattice distortion wave can tunnel through a thin insulator barrier, even though the total energy of the electron is lower than the potential height of the thin insulator barrier.

Under an electromagnetic field with a frequency $f'$, the graph of I - V characteristic shows an ohmic behavior with the constant - voltage steps in which the voltage spacing of the steps is $\Delta V = hf'/2e$ or $f' = 2e\Delta V/h$. Historically, It was just this method that provides the convincing evidence of the correctness of Josephson's predictions.[60,61] In fact, the $2\Delta V$ in the equation for frequency $f'$ is also resulted from the voltage drops caused by radiation field across two superconductor sections on both sides of a junction.

Another cornerstone for the BCS theory is the unit of a magnetic flux quantization, $h/2e$, which leads to a conclusion that the Cooper pair current is induced in a superconducting ring. In fact, the supercurrent induced in a type 1 superconducting ring consists of two equal currents, which are induced by the twice changes of magnetic flux in the ring. Now let us discuss why the double equal currents can be induced at the inner surface of a type I superconducting ring.

When a solid cylinder made of a type I superconductor is cooled into the superconducting state in the presence of a magnetic field, the magnetic flux is expelled from the interior of the cylinder by the circulating current formed at the surface of the cylinder. This phenomenon is known as the Meissner effect. Now let us assume that the same type I superconductor made in the shape of a hollow cylinder is placed in an external magnetic field with its axis parallel to the flux lines. When the temperature is greater than $T_c$, the field lines penetrate the interior of the hollow cylinder and pass through the hole of the hollow cylinder. As the temperature is lowered below $T_c$, like in the solid cylinder case, the larger circulating current will be first induced at the outer surface of the hollow cylinder to expel all external magnetic flux under the cross section of the hollow cylinder. At the same time, a circulating current, which has a direction opposite to that induced at the outer surface will be induced at the inner surface of the hollow cylinder to maintain the number of the flux lines in the hole constant. Under this circumstance, the initially induced circulating current at the outer surface of the hollow cylinder would be reduced to a value by which the magnetic field produced in the interior of a hollow cylinder exactly cancels the applied magnetic field, as shown in Fig. 8(a).

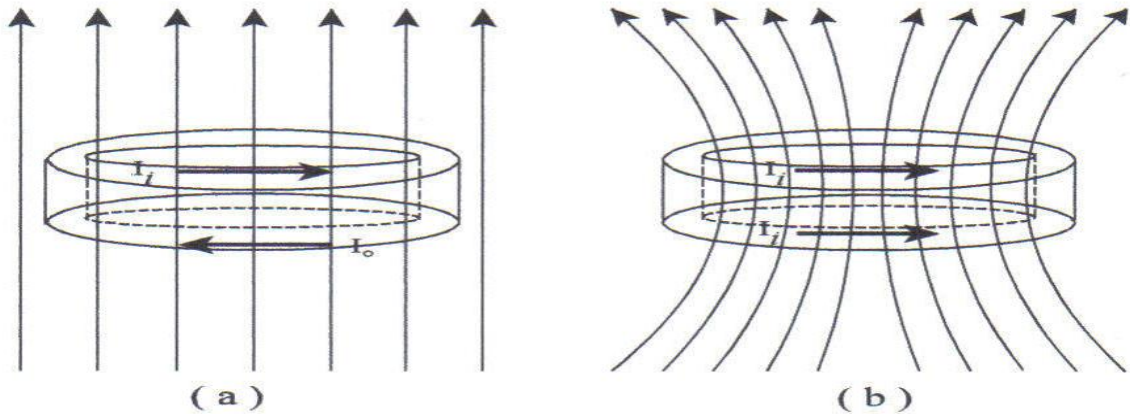

FIG. 8. Magnetic flux quantized in units of h/e in a superconducting hollow cylinder. When a type 1 superconducting hollow cylinder is cooled into superconducting state in a magnetic field, a circulating current at the outer surface of the hollow cylinder will be first induced to expel all external magnetic flux under the cross section of the hollow cylinder. At the same time, a circulating current, $I_i$, which has a direction opposite to that induced at the outer surface must be induced at the inner surface of the hollow cylinder to maintain the flux lines in the hole constant. At this time, the initially induced circulating current at the outer surface of the hollow cylinder would be reduced to a current $I_0$ with a magnitude by which the magnetic field produced in the interior of the cylinder exactly cancels the external magnetic field (a). When the external magnetic field is turned off, another circulating current with the same magnitude but in a direction opposite to the original one will be induced at the outer surface of the hollow cylinder, which will exactly cancel the original $I_0$. At the same time, an additional circulating current with the same value and direction as the first, $I_i$, would be induced at the inner surface, so that the number of flux lines passing through the hole in this case should be twice that occurred in the former case, as shown in (b).

This process explains why when a type I superconducting ring is cooled into the superconducting state in a magnetic field, the magnetic flux is expelled from the interior of the ring, but still passes through the hole of the ring. Since the supercurrent flows only in a thin penetration region in a type I superconducting ring, as long as the wall of the hollow cylinder has an enough thickness compared with the twice penetration depths, the argument above should be held. However, when the external magnetic field is turned off, another circulating current with a direction opposite to the original one will be induced at the outer surface of the hollow cylinder, which will exactly cancel the original one. That is why one cannot find any current at the outer surface of a type superconductor ring after the external magnetic field is removed. While to counteract the decrease in flux in the hole, another circulating current must be induced at the inner surface, which should have the same value and direction as the first one. According to Faraday's law, the double induced circulating current must make the total magnetic flux lines threading the hole of the hollow cylinder have a value as high as twice the original one, as shown in Fig. 8(b).

The basic nature of Faraday's law is that an electric current can be produced in a conducting loop by changing magnetic flux through the loop. This original statement of Faraday's law can be applied to any materials including superconductors. But the generalized statement about Faraday's law, which is customarily represented as that any change in magnetic flux must induce a voltage around the loop, definitely cannot be applied to superconductors. So that when a superconducting ring is cooled into its superconducting state in a magnetic field, the magnetic flux maintained in the hole is not due to the zero voltage dropped across the ring, but resulting from the circulating current induced at the inner surface of the ring. Thus we arrive at a conclusion that after the external magnetic field is removed, there exist two equal circulating currents at the inner surface of a superconducting ring, which is induced by the twice changes of magnetic flux in the hole of the ring [see Fig. 8(b)]. We cannot attribute these two equal currents induced at the inner surface of a ring to the Cooper current. Also, after the external magnetic field is removed, the doubling magnetic flux threading the hole of a superconducting ring cannot derive the conclusion that the magnetic flux in a superconducting ring is quantized in the units of $h/2e$. If there does exist the quantization of the magnetic flux in a superconducting ring, then the magnetic flux should be quantized in units of $h/e$, the original one postulated by London.

## VI. CONCLUSION AND DISCUSSION

As is well known, the most fundamental law in the natural world perhaps is of energy conservation. That is, the energy in an isolated system can be transformed from one form to another, but the total energy in the system must be always conserved. So we can expect that only the theory that is built on the basis of the total energy of a system can provide a unified explanation of a wide variety of the phenomena due to the energy transform inside the system. However, the energy carried by the nonbonding electrons in open – shell compounds has been ignored in the fundamentals of condensed - matter physics, so that all phenomena related to this kind of energy, such as lattice distortions, anharmonic off – site vibrations, phase transitions, and superconductivity, etc all become big puzzles. Once we take account of the CIDSE caused by the high – energy nonbonding electrons in open – shell compounds, then all phenomena above can be consistently understood.

Based on carrier – induced dynamic strain effect, we propose that superconducting state consists of the dynamic bound state of superconducting electrons, which is formed by the high – energy nonbonding electron through dynamic interaction with its surrounding lattice to trap itself into the three–dimensional potential well lying in energy at above the Fermi level of the material. The concept of dynamic bound state defined in this paper is totally different from the conventional bound states defined in solid state physics, which are all originated from normal electrons through electrostatic interaction with lattice. Based on the discussions about the superconductivity mechanism and its application in cuprates, we conclude that this dynamic bound state model does hold the nature of superconductivity. Almost all of the physical properties of cuprates observed in both their normal and superconducting states can be consistently explained by this superconductivity mechanism. In addition, the central features of superconductivity, like Josephson effect, the origin of superconducting tunneling phenomenon, the tunneling mechanism in multijunction systems, as well as the unit of magnetic flux quantization are all physically reconsidered under this new model.

According to this superconductivity model, the binding energy of superconducting electrons in their dynamic bound state dominates the superconducting transition temperature of the material. Under an electric field, the superconducting electrons move coherently with lattice distortion wave and periodically exchange their excitation energy with chain lattice, that is, the superconducting electrons transfer periodically between their dynamic bound state and conducting state. So the superconducting electrons cannot be scattered by chain lattice, and the supercurrent persists in time. Thus, the intrinsic feature of superconductivity is to generate an oscillating current under a dc voltage. The wave length of an oscillating current equals the coherence length of superconducting electrons along the chain direction. The coherence lengths in cuprates must have the values equal to an even number times the lattice constant, while as coherence lengths equal an odd number times the lattice constant, the superconducting process cannot be started. A superconducting material must simultaneously satisfy the following three necessary conditions required by superconductivity.

First, the material must possess the high-energy nonbonding electrons with certain concentrations requested by coherence lengths. Following this criterion, it is not surprising that most of alkaline metal, the covalent and closed-shell compounds, and the excellent conductors, copper, silver and gold do not show superconductivity at normal condition.

Second, the material must have the three-dimensional potential wells lying in energy at above the Fermi level of the material, and the dynamic bound state of superconducting electrons in potential wells of a given superconducting chain must have the same binding energy and symmetry. According to the types of potential wells in which the superconducting electrons trap themselves to form superconducting dynamic bound state, the superconductors as a whole can be divided into two groups. One of them is called as usual as the conventional superconductors in which the potential well are formed by the microstructures of materials, such as crystal grains, clusters, nanocrystals, superlattice, and the charge inversion layer in metal surfaces. We propose that the type 1 superconductors are most likely achieved by the last kind of potential wells above. The common feature for this sort of superconductors is that the volume of the potential wells for trapping superconducting electrons varies with the techniques using to synthesize the superconductors, so that the superconducting transition temperature in conventional superconductors usually shows strongly sample-dependent and irreproducible. Since the potential wells in conventional superconductors generally have relatively large confined volume and low potential height, so the conventional superconductors normally have relatively low transition temperature, but magnesium diboride is an exception. Another group is referred to as the high-$T_c$ superconductors in which the potential wells for trapping superconducting electrons are formed by the lattice structure of material only, such as $CuO_6$ octahedrons and $CuO_5$ pyramids potential wells for cuprates, $BiO_6$ octahedron for $BaKBiO_3$ compounds, $C_{60}$ in $A_3C_{60}$ fullerides and $FeAs_4$ tetrahedrons in LaOFeAs compounds. The small and fixed volume of potential wells makes the high-$T_c$ superconductors usually have relatively high and fixed transition temperature.

Finally, in order to enable the normal state of the material being metallic, the band structure of the superconducting material must have a widely dispersive antibonding band, which crosses the Fermi level and runs over the height of potential wells. The symmetry of the antibonding band into which the superconducting electrons trap themselves to form a dynamic bound state dominates the types of the superconducting distortion waves. The typical example for superconductivity derived from this criterion perhaps belongs to transition metals and their compounds. Matthias was the first to propose that the transition temperature in transition metals depends upon the number of valence electrons per atom, Ne, and two values Ne = 5e/a for V, Nb, and Ne = 7e/a for Tc and Re are favorable to have high value of $T_c$.[62] The similar phenomenon was also found in transition metal compounds. It has been confirmed that the density of electronic states for both bcc and hcp transition metals are all resulted from a number of the narrow density peaks derived from the d - orbitals bonding states overlapping with a broad low density of states arisen from the s - electron antibonding band. Based on the rigid band model, the Fermi levels for the transition metal with Ne = 1 to 4 all fall in the region where the density of states is dominated by the d - electron bonding states. The potential wells formed by the grain boundaries, which normally have a potential height less than 0.1 eV, should also overlap with bonding states of the d - orbitals. In this case, the dynamic bound state cannot be formed in the potential wells, thus it is not surprising that the superconductivity cannot be found in these transition metals. However, for V and Nb, which have five valence electrons, Ne = 5e/a, the Fermi level shifts toward the high energies at where the density of states is mainly resulted from the s electron-antibonding band. In this circumstance, the energy levels at the top of potential wells formed by grain boundaries are derived from the s electron-antibonding band, and so the superconducting state can be achieved and has a s-symmetry wave. The similar process is repeated for the transition metal Tc and Re with Ne = 7 e/a (Ref. 62).

On the basis of the mechanism of superconductivity proposed above, the key point to achieve superconductivity is that the superconducting electron must periodically exchange its excitation energy with chain lattice. That is, the excitation energy of the superconducting electrons must be reversibly transferred between superconducting electrons and chain lattice. It is well known that the interaction between electrons and atomic magnetic moments is irreversible, which, thus, in any case cannot become the driving force of superconductivity. However, it can be seen from this new model that superconductivity and atomic magnetic moments in principle are not intrinsically exclusive each other. As long as there exists the same magnetic moment in every potential well in a given superconducting chain, as in the case of the ferromagnetic materials LaOFeAs, and the three necessary conditions required for superconductivity are satisfied, the superconducting state can be formed and the superconducting process will persist without dissipating energy. Since the electromagnetic interaction energy for superconducting electrons with atom magnetic moment maintains the same in every potential well, thus the binding energy of superconducting electrons in potential wells cannot be affected by the atom magnetic moment, and so the scattering centers for superconducting electrons cannot be introduced. But this condition essentially cannot be achieved for conventional superconductors, so the atomic magnetic moments are generally detrimental to superconductivity.

The possibility to achieve the room temperatures superconductivity has been argued for decades in the superconductivity research field. Because the real mechanism of superconductivity has never been revealed, so the estimates about the upper bound on the superconducting transition temperature are all empirical. Based on the superconductivity mechanism proposed in this paper, clearly, the room temperatures superconductivity must lie in the materials in which the three criteria for superconductivity have to be optimally satisfied. For the time being, we cannot predict what the upper bound of the superconducting transition temperature should be, but we assert that it is definitely higher than the room temperatures. We believe that the dream to achieve the room temperatures superconductivity will come true in the near future.

ACKNOWLEDGMENT

Superconductivity, since it was discovered, has brought great hope and puzzle together into the physical world, while to find the real mechanism of the superconductivity has become a historical task of the entire superconductivity community all over the world. For this goal, thousands upon thousands of scientists have continuously worked in this field for almost a century. In this cooperative effort, everyone is the gainer and winner, we all have benefitted from what they have done. Here we would like to deeply thank the pioneers for their outstanding contributions which have laid down a solid foundation for further understanding the nature of superconductivity. Specially, we would like to thank W. E. Pickett for his excellent review article entitled "Electronic structure of the high- temperature oxide superconductors ", which provides us with a highly effective understanding about the physical properties of cuprates.[2] In addition, we have to apologize for neglecting the great number of excellent works which may also strongly support the model we proposed. Particularly, we have to thank God for giving us an opportunity to work at the physical mechanism of superconductivity. We would like to thank professor Xi-Cheng Zhang in department of physics at Rensselaer Polytechnic Institute for helpful discussions.